%% file: GalMock.tex
\documentclass[usenatbib]{mn2e}
\usepackage{graphicx}
\usepackage{natbib}
\usepackage{amsmath}
\include{aas_journals}

\title[Mock galaxy catalogues]
{Mock galaxy redshift catalogues from simulations: implications for Pan-STARRS1}

\begin{document}

\author[Cai et al.]{
\parbox[h]{\textwidth}
{Yan-Chuan Cai, Raul E. Angulo, Carlton M. Baugh, Shaun Cole, Carlos
  S. Frenk, Adrian Jenkins} \vspace*{6pt} \\
Institute for Computational Cosmology, Department of Physics, University of Durham, South Road, Durham, DH1 3LE, UK}
\maketitle

\begin{abstract}

We describe a method for constructing mock galaxy catalogues which are
well suited for use in conjunction with large photometric surveys.
We use the semi-analytic galaxy formation model of Bower et al. 
implemented in the Millennium N-body simulation of the evolution 
of dark matter clustering in a $\Lambda$CDM cosmology. 
We apply our method to the specific case of the surveys
soon to commence with PS1, the first of 4 telescopes planned for the
Pan-STARRS system. PS1 has 5 photometric bands, $g, r, i, z$ and $y$
and will carry out an all-sky ``3$\pi$'' survey and a medium 
deep survey (MDS) over 84 sq. deg. 
We calculate the expected magnitude limits for extended sources 
in the two surveys. We find that, after 3 years, the 3$\pi$ survey 
will have detected over $10^8$ galaxies in all 5 bands, 10 million 
of which will lie at redshift $z>0.9$, while the MDS will have detected 
over $10^7$ galaxies with 0.5 million lying at $z>2$. 
These numbers at least double if detection in the shallowest band, $y$ 
is not required. 
We then evaluate the accuracy of photometric redshifts estimated 
using an off-the-shelf photo-$z$ code. 
With the $grizy$ bands alone it is possible to
achieve an accuracy in the 3$\pi$ survey of $\Delta z/(1+z)\sim 0.06$
in the range $0.25<z<0.8$, which could be reduced by about 15\% using
near infrared photometry from the UKIDDS survey, but would increase by
about 25\% for the deeper sample without the $y$ band photometry. For 
the MDS an accuracy of $\Delta z/(1+z)\sim 0.05$ is achievable for 
$0.02<z<1.5$ using $grizy$. 
A dramatic improvement in accuracy is possible by selecting
only red galaxies. In this case, $\Delta z/(1+z) \sim 0.02-0.04$ is
achievable for $\sim$100~million galaxies at $0.4<z<1.1$ in the 3$\pi$
survey and for 30~million galaxies in the MDS at $0.4<z<2$. 
We investigate the effect of using photometric redshifts in the 
estimate of the baryonic acoustic oscillation scale. We find that PS1 
will achieve a similar accuracy in this estimate as a spectroscopic 
survey of 20 million galaxies. 
\end{abstract}

 
%===================================================================
 
\section{Introduction}

Studies of the cosmic large structure were brought to a new level 
by the two large galaxy surveys of the past decade, the
``2-degree-field galaxy redshift survey \citep[2dFGRS;][]{Colless01}
and the Sloan Digital Sky Survey \citep[SDSS][]{York00}.
The former relied on
photographic plates for its source catalogue while the latter was
compiled from the largest CCD based photometric survey to date. Most of the
large-scale structure studies carried out with these surveys made use
of spectroscopic redshifts, for about 220,000 galaxies in the case of
the 2dFGRS and 585,719 galaxies in the case of the SDSS 
\citep{Strauss02}. These surveys achieved
important advances such as the confirmation of the existence of dark
energy \citep{Efstathiou02,Tegmark04}
and the discovery
of baryonic acoustic oscillations \citep{Percival01,Cole05, Eisenstein05}.
Yet, a number of fundamental
questions on the cosmic large-scale structure remain unanswered, such
as the identity of the dark matter and the nature of the dark energy.

Further progress in the subject is likely to require a new generation
of galaxy surveys at least one order of magnitude larger than the
2dFGRS and the SDSS. Unfortunately, measuring redshifts for millions
of galaxies is infeasible with current instrumentation. Attention has
therefore shifted to the possibility of carrying out extremely large
surveys of galaxies in which, instead of using spectroscopy, redshifts
are estimated from deep multi-band photometry. Although the accuracy
of these estimates is limited, this strategy can yield measurements
for hundreds of millions of galaxies or more. Several instruments are
currently being planned to carry out such a programme. The most
advanced is the Panoramic Survey Telescope \& Rapid Response System
\citep{Chambers06}.
Of an eventual 4 telescopes for this
system, the first one, PS1, is now in its final commissioning stages
and is expected to begin surveying the sky early in 2009. This
telescope is likely to be quickly followed by the full Pan-STARRS
system and by the Large Synoptic Survey Telescope \citep[LSST][]{Tyson02}.
Several other smaller photometric surveys are currently
underway \citep[UKIDSS, Megacam etc.][]{Lawrence07,Boulade03}.

One of the important lessons learned from previous surveys, including
2dFGRS and SDSS, is the paramount importance of careful modelling of
the survey data for the extraction of robust astrophysical
results. Such modelling is best achieved using large cosmological
simulations to follow the growth of structure in a specified
cosmological background. The simulations can be used to create mock
versions of the real survey in which the geometry and selection
effects are reproduced. Such mock surveys allow a rigorous assessment
of statistical and systematic errors, aid in the design of new
statistical analyses and enable the survey results to be directly
related to cosmological theory. Mock catalogues based on cosmological
simulations were first used in the 1980s, in connection with the CfA
galaxy survey \citep{Davis85, White88} and redshift
surveys of IRAS galaxies \citep{Saunders91} and have been
extensively deployed for analyses of the 2dFGRS and the SDSS \citep{Cole98,
Blaizot06}.

The recent determination of the values of the cosmological parameters
by a combination of microwave background and large-scale structure
data \citep[e.g.][]{Sanchez06,Komatsu08} has removed one major layer
of uncertainty in the execution of cosmological simulations. N-body
techniques are now sufficiently sophisticated that the evolution of
the dark matter can be followed with impressive precision from the
epoch of recombination to the present \citep{Springel05}. The main
uncertainty lies in calculating the evolution of the baryonic
component of the Universe.

The size of the planned photometric surveys and the need to understand
and quantify uncertainties in estimates of photometric redshifts and
their consequences for diagnostics of large-scale structure pose novel
challenges for the construction of mock surveys. The simulations need
to be large enough to emulate the huge volumes that will be surveyed
and, at the same time, the modelling of the galaxy population needs to
be sufficiently realistic to allow an assessment of the uncertainties
introduced by photometric redshifts. At present, the only technique
that can satisfy both these two requirements is the combination of
large N-body simulations with semi-analytic modelling of galaxy
formation.

Semi-analytic models of galaxy formation are able to follow the
evolution of the baryonic component in a cosmological volume by making
a number of simplifying assumptions, most notably that gas cooling
into halos can be calculated in a spherically symmetric approximation
\citep{White91}. Once the gas has cooled, these models employ
simple physically based rules, akin to those used in hydrodynamic
simulations, to model star formation and evolution and a variety of
feedback processes. The analytical nature of these models makes it
possible to investigate galaxy formation in large volumes and to
include, in a controlled fashion, a variety of processes, such as dust
absorption and emission, that are currently beyond the reach of
hydrodynamic simulations. \citep[For a review of this approach, 
see][]{Baugh06}. It is reassuring that the simplified
treatment of gas cooling in these models agrees remarkably well with
the results of full hydrodynamic simulations \citep{Helly03,Yoshida02}.

An important feature of semi-analytic models is that they are able to
reproduce the local galaxy luminosity function from first principles 
\citep[e.g.][]{Cole00, Benson03, Hatton03, 
Baugh05,Kang05,Kang06, Bower06,Croton06,deLucia06} and, in the most
recent models, also its evolution to high redshift
\citep{Bower06,deLucia06,Lacey08}. These recent models also provide a
good match to the distribution of galaxy colours, which is
particularly relevant for problems relating to photometric
redshifts. And, of course, the models also calculate many properties
which are not directly observable (e.g. rest-frame fluxes, stellar
masses, etc) but which are important for the interpretation of the
data.

There are currently two main approaches to the estimation of
photometric redshifts. One employs an empirical relation, obtained by
fitting a polynomial or a more general function derived by an
artificial neural network, between redshift and observed properties,
such as fluxes in specified passbands
\citep[e.g.][]{Connolly95,Brunner00,
Sowards-Emmerd00, Firth03, Collister04}. The
second method is based on fitting the observed spectral energy
distribution (SED) with a set of galaxy templates 
\citep[e.g.][]{Sawicki97, Giallongo98,
Bozonella00,Benitez00, Bender01, Csabai03},
obtained either from observations of the local universe
\citep[e.g.][]{Coleman80} or from synthetic spectra
\citep[e.g.][]{Bruzual93,Bruzual03}.
Some authors \citep[e.g.][]{Collister04} claim that the empirical 
fitting method can give smaller redshift errors, but this method relies on 
having a well-matched spectroscopic subsample that reaches the same depth in 
every band as the photometric survey. Unfortunately, for Pan-STARRS
or LSST this is going to be challenging and to be conservative in this
initial investigation we use Hyper-$z$ because it does not
require a training set.

In this paper, we describe a method for generating mock catalogues
suitable, in principle, for the next generation of large photometric
surveys. As an example, we construct mocks surveys tailored to
PS1. PS1 will carry out two, 3-year long surveys in 5 bands
($g,r,i,z,y$), the ``$3\pi$ survey'' which will cover three quarters
of the sky to a depth of about $r=24.5$ and the ``medium deep survey''
(MDS) which will cover $84$ sq deg to a $5-\sigma$ point sources 
depth of $r=27$ (AB system). The former will enable a comprehensive 
list of large-scale structure
measurements, including the integrated Sachs-Wolfe effect and baryonic
acoustic oscillations. The latter will be used to study clustering on
small and intermediate scales, as well as galaxy evolution. To
construct the mocks we use the semi-analytic model of
\citet[hereafter B06]{Bower06} as implemented in the Millennium
simulation \citep{Springel05}. 
A sophisticated adaptive template method based on the work of 
\citet{Bender01} will be employed for the genuine PS1 survey.
The method has been applied in the photo-$z$ measurements of FORS Deep Field 
galaxies \citep{Gabasch04} and achieved $\Delta z/(1+z_{\rm{spec}}) \leq 0.03$ 
with only $1\%$ outliers. However, the method requires precise 
calibration of zeropoints in all filters using the colour-colour plots 
of stars, and a control sample of spectroscopic
redshifts. Consequently this method cannot be rigorously tested until
genuine PS1 data is available. Therefore, for a first look at 
the photo-$z$ performance of PS1, we adopt the standard SED fitting 
method as implemented in the Hyper-$z$ for our mock catalogues.

The paper is organised as follows. In \S2, we briefly summarise the
models and detail the process of constructing mock galaxy
catalogues. In \S3 we analyse some of their properties and in \S4 we
use the mock catalogues to assess the accuracy with with photometric
redshifts will be estimated by PS1. In \S5, we discuss how these
uncertainties are likely to affect the accuracy with which baryonic
acoustic oscillations, one the main targets for PS1, can be measured
in the survey. Finally, in \S6, we discuss our results and present our
conclusions.

%==========================================================================

\section{Mock catalogue construction}

In this section we describe how we construct mock catalogues. In
\S2.1 we describe the semi-analytic galaxy formation code that we
use. In \S2.2 we compare the luminosities and sizes of the model
galaxies to SDSS data and modify them to improve the accuracy of the
mocks.  Finally in \S2.3 we describe how the mock catalogues themselves
are built.

\subsection{The galaxy formation model} 

\begin{figure*}
\begin{center}
\includegraphics[width=8.5cm,angle=0]{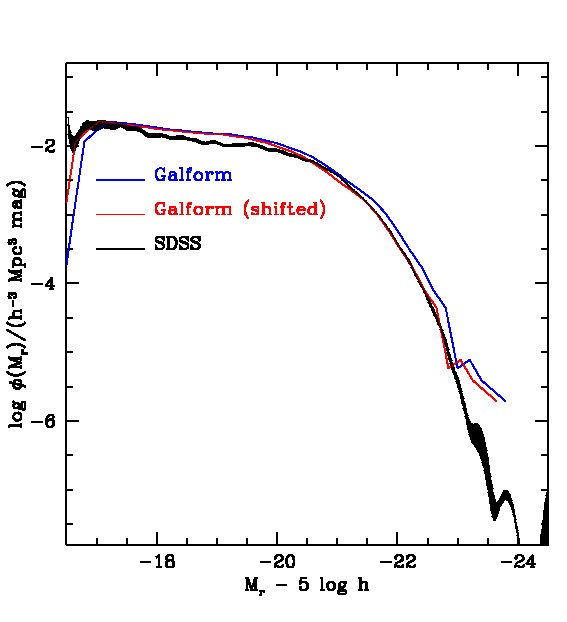}
\includegraphics[width=8.5cm,angle=0]{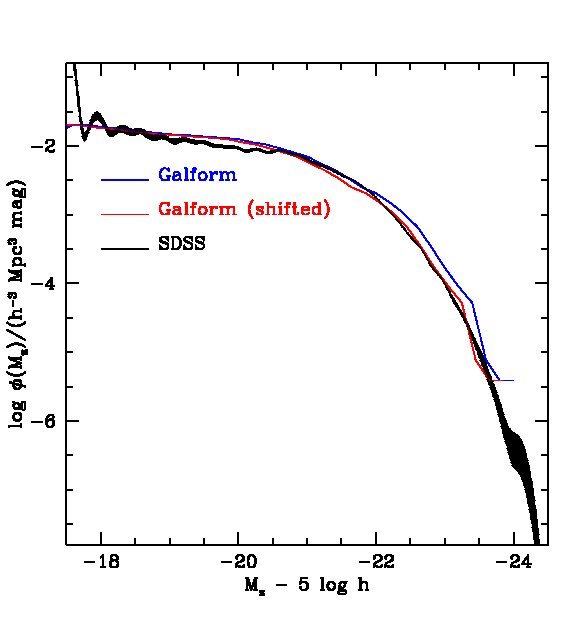}
\caption{Luminosity functions predicted by {\sc galform}, compared 
with the SDSS results in the $r$-band (left) and $z$-band
(right), table from \citet{Blanton03}. 
The black lines with error bars indicated by the shaded region
are the SDSS results. The
blue lines show the original {\sc galform} prediction, while the red
lines show the {\sc galform} prediction globally shifted faintwards by 0.15
magnitudes.}
\label{figLF1}
\end{center}
\end{figure*} 
The first step in the process of generating a mock catalogue is to
produce a population of model galaxies over the required redshift
range.  We use the {\sc galform} semi-analytic model of galaxy
formation \citep{Cole00,Benson03,
Baugh05, Bower06} to do this.  {\sc galform} calculates
the key processes involved in galaxy formation: (i) the growth of dark
matter halos by accretion and mergers; (ii) radiative cooling of gas
within halos; (iii) star formation and associated feedback processes
due to supernova explosions and stellar winds; (iv) the suppression of
gas cooling in halos with quasistatic hot atmospheres and accretion
driven feedback from supermassive black holes (see
\citet{Malbon06} for a description of the model of black
hole growth); (v) galaxy mergers and the associated bursts of star
formation; (vi) the chemical evolution of the hot and cold gas, and
the stars.

{\sc galform} uses physically motivated recipes to model these
processes. Due to the complex nature of many of them, the model
necessarily contains parameters which are set by requiring that it
should reproduce a subset of properties of the observed galaxy
population \citep[see][for a discussion of the philosophy behind
setting the values of the model parameters]{Cole00, Baugh06}. 

{\sc galform} predicts star formation histories for the population of
galaxies at any specified redshift. These histories are far more
complicated than the simple, exponentially decaying star formation
laws sometimes assumed in the literature \citep[for examples of star
formation histories of {\sc galform} galaxies, see][]{Baugh06}. 
The {\sc galform} histories have the
advantage that they are produced using an astrophysical model in
which the supply of gas available for star formation is set by 
source and sink processes. The sources are the infall of new material due
to gas cooling and galaxy mergers and gas recycling from previous
generations of stars. The sinks are star formation and the
reheating or removal of cooled gas by feedback processes. The
metallicity of the gas consumed in star formation is modelled using
the instantaneous recycling approximation and by following the
transfer of metals between the hot and cold gas, and the stellar reservoirs
\citep[see Fig. 3 of ][]{Cole00}. 

The model outputs the broad band magnitudes of each galaxy for a set
of specified filters. In this paper we use the PS1 filter set ($g$,
$r$, $i$, $z$, $y$) \citep{Chambers06}, augmented by a few additional filters where some
of the PS1 galaxies may be observed as part of other observational
programmes ($U$, $B$, $J$, $H$, $K$). In addition to the magnitudes of
model galaxies in the observer's frame, we also output the rest frame
$g-r$ colour, in order to distinguish between red and blue galaxy
populations. {\sc galform} tracks the bulge and disk components of the
galaxies separately \citep{Baugh96}. The scale sizes of
the disk and bulge (assumed to follow an exponential profile and an
$r^{1/4}$-law in projection, respectively) are also calculated
\citep{Cole00} \citep[see][for a test of the prescription for 
computing the size of the spheroid component]{Almeida07}. 

The B06 model which we use in this paper employs halo merger trees
extracted from the Millennium N-body simulation of a $\Lambda$-cold
dark matter universe \citep{Springel05}. The model gives a very good
reproduction of the shape of the present day galaxy luminosity
function in the optical and near-infrared. Also of particular
relevance for the predictions presented here is the fact that this 
model matches the observed evolution of the galaxy
luminosity function.

\subsection{Improving the match to SDSS observations}

To make the mocks as realistic as possible, we modify the luminosities
and sizes of the model galaxies to give a better fit to SDSS
data. Although both the $b_J$-band and K-band luminosity functions of
the B06 model have been shown to agree well with observations of the local
universe, the agreement is not perfect and a shift of $0.15$ magnitude
faintwards in all bands improves the match to the data as can
be seen in Fig.~\ref{figLF1}.  The original B06 K-band luminosity
functions also match observations up to redshift $z=1.5$
\citep{Bower06}, although the observational error bars are relatively
large.  Hence, even after applying the 0.15~magnitude shift the
agreement between model and high redshift observations remains reasonably good.

The {\sc galform} magnitudes we have been dealing with so far are
total integrated magnitudes.  In reality all but the most distant
galaxies in the survey will be resolved over several pixels and will
have lower signal to noise in each of these pixels than a point source
would. To take this into account it proves convenient to use
\cite{Petrosian76} magnitudes. These have the advantage over fixed
aperture magnitudes that, for a given luminosity profile shape, they
measure a fixed fraction of the total luminosity independently of the
angular size and surface brightness of the galaxy. The Petrosian flux 
within $N_p$ times the
Petrosian radius, $r_P$, is:
\begin{equation}\label{eq2}
F_P=2\pi\int^{N_Pr_P}_{0}I(r)r\rm{d}r,
\end{equation}
where $I(r)$ is the surface brightness profile of the galaxy. The
Petrosian radius is defined such that at this radius, the ratio of the
local surface brightness in an annulus at $r_P$ to the mean surface
brightness within $r_P$, is equal to some constant value $\eta$,
specifically:
\begin{equation}\label{eq3}
\eta=\frac{2\pi
\int^{1.25r_P}_{0.8r_P}I(r)\rm{d}r/[\pi((1.25r_P)^2-(0.8r_P)^2)]}{2\pi
\int^{r_P}_0I(r)r\rm{d}r/(\pi r^2_P)}.
\end{equation}
We choose the parameter values as $N_P=2$ and $\eta=0.2$ as adopted
in the SDSS \citep[e.g.][]{Yasuda01}.

We decompose the surface brightness profile of each galaxy, $I(r)$,
into the superposition of a disk and a bulge: $I(r)=I_{\rm disk}(r)+I_{\rm
bulge}(r)$.  The disk component is taken to have a pure exponential
profile:
\begin{equation}\label{eq4}
I_{\rm disk}(r)=I_0e^{-1.68r/r_d},
\end{equation}
and the bulge a pure de Vaucouleurs profile :
\begin{equation}\label{eq5}
I_{\rm bulge}(r)=I_0e^{-7.67[(r/r_b)^{1/4}]}.
\end{equation}
Given these assumptions, and assuming the disks are face-on, 
we can compute the Petrosian radius by solving
Eqn.~(\ref{eq3}) for each {\sc galform} galaxy.
 
While {\sc galform} does provide an estimate of the disk and bulge
sizes, it has been shown by \citet{Almeida07} that the early
type galaxy sizes of the B06 model are not in particularly good
agreement with the SDSS observational results. Therefore, for the
purposes of producing more realistic mocks, we modify the galaxy sizes
so as to match the SDSS results given by \citet{Shen03}.

To do this we need to separate the galaxies into early and late types
and apply separate corrections to each population. First, we use the
concentration parameter defined as $C=R_{90}/R_{50}$ to separate our
galaxies into early and late types, where $R_{90}$ and $R_{50}$ are
the Petrosian $90\%$ and $50\%$ light radii respectively.  We then
calculate the ratio of the {\sc galform} galaxy size to the mean found
by \citet{Shen03} for SDSS galaxies as a function of galaxy magnitude 
and obtain the average
correction factor as a function of magnitude required for the {\sc
galform} galaxies to match the SDSS size data.  For early type
galaxies at redshift $z=0.1$ 
\citet{Shen03} parameterised the relation between
Petrosian half-light radii in the $r$-band, $R_{50}$, and absolute
$r$-band magnitude, $M$, as
\begin{equation}
\log(R_{50})=-0.4aM+b, 
\end{equation}
with $a=0.60$ and $b=-4.63$. While for late type galaxies they found
\begin{equation}
\log(R_{50})=-0.4\alpha
M+(\beta-\alpha)\log[1+10^{-0.4(M-M_0)}]+\gamma, 
\end{equation}
with $\alpha=0.21$, $\beta=0.53$, $\gamma=-1.31$ and $M_0=-20.52$.
To correct the {\sc galform} galaxy sizes at other redshifts
we adopt $R_{50} \propto 1-0.27z$ for late type galaxies and $R_{50}
\propto -0.33z+1.03$ for  early type galaxies. The former is the
relation given by \citet{Bouwens02} which agrees with a
combination of SDSS, GEMS and FIRES survey data
\citep{Trujillo06}. The relation for early type
galaxies is obtained by taking a linear fit to the data given by
\citet{Trujillo06}. Finally, we apply a linear relation
between $R_{50}$ and the Petrosian radius, $R_{50}=0.47 r_{P}$.

\subsection{Building the mock catalogues}

Our goal here is to generate mock catalogues which have 
the distribution of galaxy redshifts and magnitudes expected 
for the various PS1 catalogues. For the purposes of this paper, 
we do not need to retain the clustering information contained in 
the Millennium Simulation. We are effectively generating a Monte-Carlo 
realisation of the redshift distribution expected for a given 
set of magnitude limits. The production of mock catalogues 
with clustering information will be described in a later paper. 

There are 37 discrete output epochs in the Millennium simulation
between $z=0$ and $z=3$.  The spacing of the output times is comparable 
to the typical error on the estimated value of photometric redshifts, 
as we will see later. To avoid the introduction of systematic errors 
caused by the discrete spacing of simulation output times, previous 
work to build mock catalogues used an interpolation of galaxy properties 
between output times \citep{Blaizot05}. 
We follow an alternative approach in this paper. We have 
generated nine additional outputs which are evenly spaced between 
each pair of Millennium simulation outputs. To produce {\sc galform} 
output at each of these intermediate steps, the Millennium
simulation merger trees ending at the nearest simulation output are
used but their redshifts are re-labelled to match the required
redshift. Then {\sc galform} computes the star formation history up to
the new output redshift, following the baryonic physics up to that
point. This results in a much finer spacing of effective output
redshifts which fully takes account of k-corrections, star formation
and stellar evolution, but ignores the evolution in the dark matter
distribution between the chosen output redshift and the nearest
simulation redshift.

To generate a mock catalogue with a smooth redshift distribution 
we proceed as follows. At each of our closely spaced 
grid of redshifts, $z_i$, we have a {\sc galform} output dataset 
consisting of a set of {\sc galform} galaxies sampling a fixed comoving 
volume $V_{\rm GF}$ down to a sufficiently deep absolute magnitude. 
To each of these
datasets we apply a magnitude limit and record the number of galaxies,
$N_i$, that remain.  The comoving number density of galaxies expected
brighter than the limit is then $n(z_i)=N_i/V_{\rm GF}$ and the number
we expect per unit redshift in the survey is $n(z)\Omega
dV/dz/d\Omega$, where $\Omega$ is the solid angle of the survey and
$dV/dz/d\Omega$ is the comoving volume per unit redshift and solid
angle for the adopted cosmology.  This can be used to compute,
$N(z_i)=\int^{z+\Delta z/2}_{z-\Delta z/2}(dN(z)/dz) dz$, the number
of galaxies expected in the survey in a redshift bin $\Delta z$,
centred at a given redshift, $z_i$. To create a continuous redshift
distribution we sample at random this number of galaxies from the
corresponding {\sc galform} output and assign them a random redshift
in the interval $\Delta z$ such that we uniformly sample the volume
redshift relation. As we have perturbed the redshift of each galaxy, we
correspondingly perturb its apparent magnitude according to the 
difference in distance modulus between the output and assigned
redshift. We will see that the residual redshift quantisation in the
evolutionary and k-corrections is small compared with the precision
achievable for the photometric redshifts. Therefore these residual
discreteness effects are not important in the photometric redshift
error estimation.

For the purpose of producing predictions for the redshift distribution 
and number counts of galaxies in PS1 surveys, and to provide an input 
catalogue with which to test photometric redshift estimators, we generate 
a mock catalogue which corresponds to a solid angle of 10 square degrees. 
We generate predictions for the $3\pi$ survey and the MDS by scaling 
the results from this mock to take into account the difference in 
solid angle. In a later paper, we will generate mock catalogues for 
clustering applications which will have the full sky coverage of these 
surveys. 

Finally we need to apply the magnitude limit. To do this, we use a
Gaussian random number generator to sample the noise level $N_r$ of
each galaxy for the specific survey under consideration. The galaxy 
source flux $S$ is also perturbed by its noise,
$S_r=S+N_r$. We apply a 5$\sigma$ cut for selection by rejecting
galaxies with signal-to-noise ratio lower than 5.

\section{PS1 mock catalogues}

In this section, we apply the methodology described in \S2 to the
specific case of the PS1 survey. We begin by calculating the magnitude
limits which we expect to be reached in the 3$\pi$ and MDS surveys for
both point and extended sources after one and three years of
observations respectively. 

\begin{table*}
\centering
\caption{Estimated PS1 3$\pi$ and Medium Deep Survey (MDS)
sensitivities. The 3$\pi$ survey will cover three quarters of the sky,
while the MDS will cover 84 sq deg of the sky in 10 separate regions. 
$m_1$ and $\mu$ are defined in section 3.1.}
\bigskip
\begin{tabular}{lcccccccc}
\hline
Filter & Bandpass & $m_1$ & $\mu$ & exposure time &  $5\sigma$   &
$5\sigma$ &  $5\sigma$ &  $5\sigma$ \\ 
       & (nm)     &  AB   &  AB     & in 1st yr (3$\pi$)    &
pt. source  & pt. source  & pt. source  & pt. source\\  
       &          &  mag  & mag/${\rm arcsec}~^2$ & sec    &  in 1st yr
(3$\pi$)    & in 3rd yr (3$\pi$)  &  in 1st yr (MDS) &  in 3rd yr
(MDS) \\ 
\hline
$g$ & 405-550  &  24.90 & 21.90 & 60$\times$ 4 & 24.04 &  24.66 & 26.72   & 27.32  \\
$r$ & 552-689  &  25.15 & 20.86 & 38$\times$ 4  & 23.50 & 24.11 & 26.36 & 26.96    \\
$i$ & 691-815  &  25.00 & 20.15 & 60$\times$ 4 & 23.39 &  24.00  & 26.32 & 26.91   \\
$z$ & 815-915  &  24.63 & 19.26 & 30$\times$ 4 & 22.37 &  22.98  & 25.69 & 26.28   \\
$y$ & 967-1024 &  23.03 & 17.98 & 30$\times$ 4 & 20.91 &  21.52  & 24.25 & 24.85   \\
\hline 
\end{tabular}
\label{tab1}
\end{table*}

\begin{table*}
\centering
\caption{Estimated UKIDSS sensitivities. All the magnitudes are in the
AB system. The Large Area Survey (LAS) aims to map about 4000 sq 
deg of the Northern sky within a few hundred nights. The Deep
Extragalactic Survey (DXS) aims to map 35 sq deg of the sky in
three separate regions.}
\bigskip
\begin{tabular}{lccccccc}
\hline
Filter & $\lambda_{\rm{eff}}$ & $m_1$ & $\mu$ & exposure time &  $5\sigma$
& exposure time &  $5\sigma$  \\ 
       & (nm)     &  AB   &  AB     &      (LAS)     &  pt. source  &
(DXS)  & pt. source  \\ 
       &          &  mag  & mag/$asec^2$ & sec    &  (LAS)   & h & (DXS)   \\
\hline 
J & 1229.7  &  23.80 & 16.80 & 40$\times$ 4 & 20.5 & 2.1 & 23.4   \\
H & 1653.3  &  24.58 & 15.48 & 40$\times$ 4  & 20.2 &-- &--     \\
K & 2196.8  &  24.36 & 15.36 & 40$\times$ 4 & 20.1 & 1.5 &22.86    \\
\hline
\end{tabular}
\label{tab2}
\end{table*}

\begin{figure}
\resizebox{\hsize}{!}{
\includegraphics{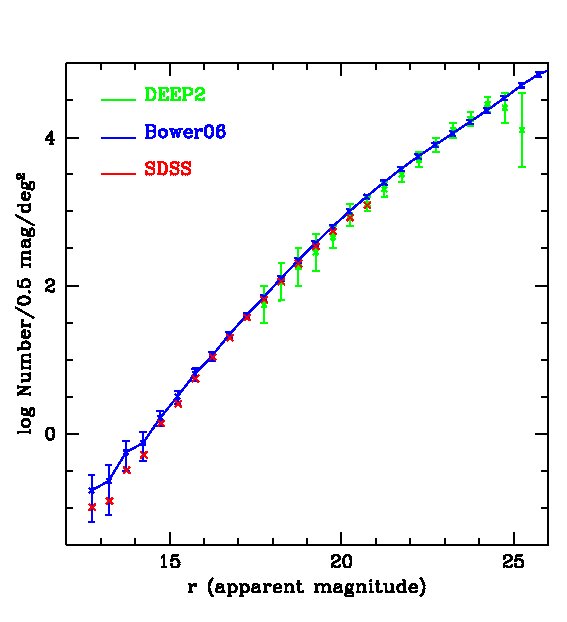}}
\caption{Galaxy number counts in 0.5 magnitude centred bins predicted
by the {\sc galform} model in the $r$-band (blue solid line with error
bars), compared with the SDSS commissioning data \citep{Yasuda01} (red crosses) and the
{\sc deep}2 survey data \citep{Coil04} (green dots with error bars). The agreement
between the model and the data is excellent. 
\label{fig2}}
\end{figure}

\begin{figure*}
\begin{center}
\includegraphics[width=8.5cm,angle=0]{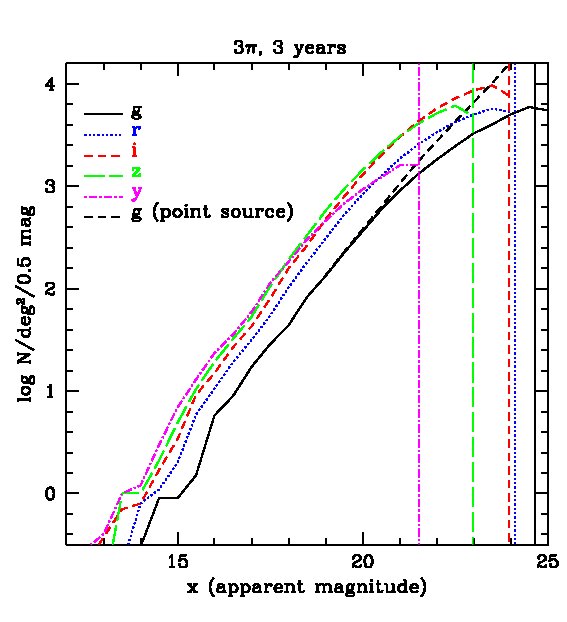}
\includegraphics[width=8.5cm,angle=0]{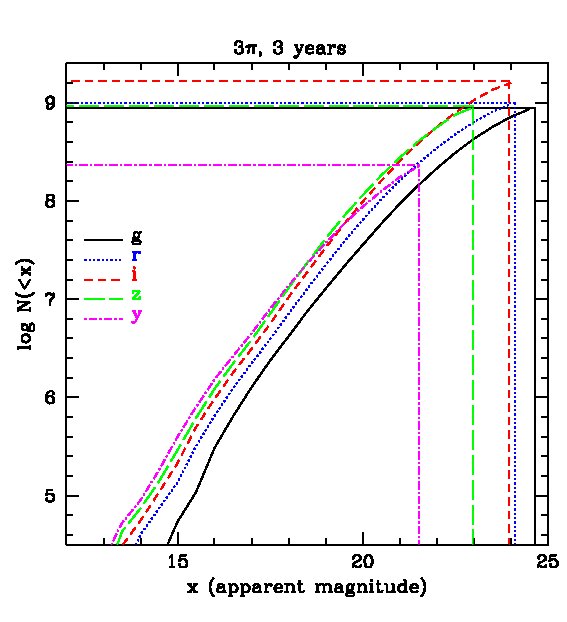}

\includegraphics[width=8.5cm,angle=0]{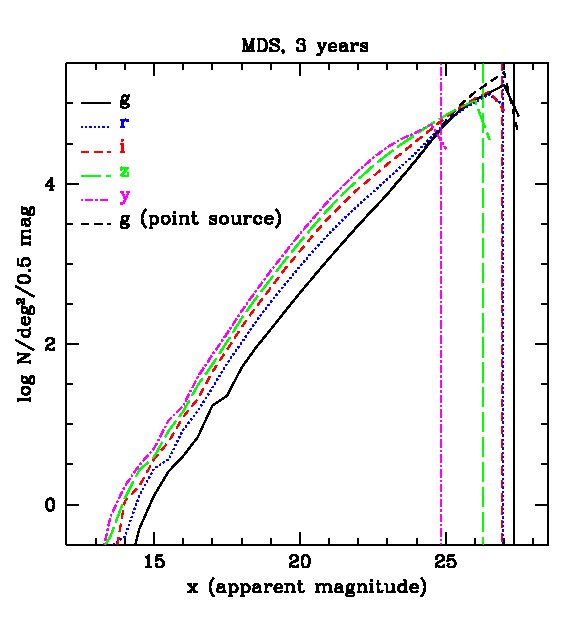}
\includegraphics[width=8.5cm,angle=0]{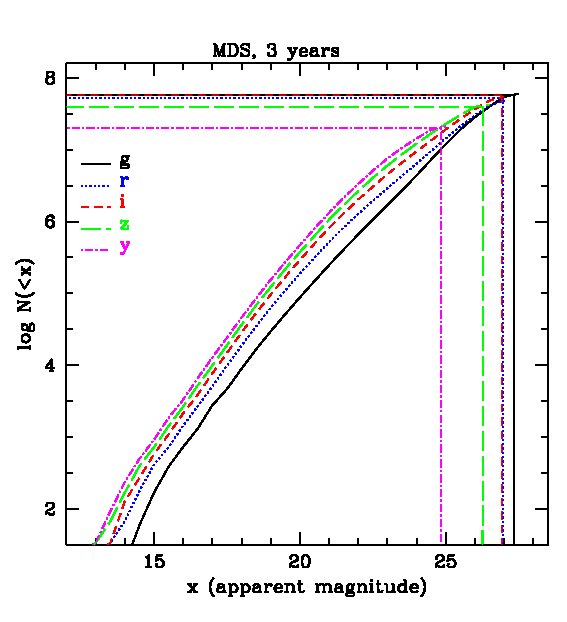}
\caption{Expected galaxy number counts in the 3-year
PS1 $3\pi$ survey (top panels) and the Medium Deep Survey (MDS)
(bottom panels), as predicted by the {\sc galform} model. A 5$\sigma$
cut on Petrosian magnitudes has been used for selecting galaxies. 
Left: galaxy number counts in 0.5 magnitude bins per sq deg 
%%in the PS1 {\it g, r, i, z}, and {\it y} bands. The dash lines show the 
%%$g$ band galaxy number counts using the point source limits. 
in the PS1 {\it g, r, i, z}, and {\it y} bands. The black dashed lines
 show the
$g$ band galaxy number counts limited only by the point source limits.
Right: cumulative galaxy number counts as a function of magnitude, 
$ N(<x)$, where $x$
%%denotes PS1 {\it g, r, i, z}, and {\it y} bands.  The straight lines
denotes PS1 {\it g, r, i, z}, or {\it y} bands, as indicated in the
legend. The straight lines
show the 3-year 5$\sigma$ point source magnitude limits. 
\label{fig5}}
\end{center}
\end{figure*}

\begin{figure*}
\begin{center}
\includegraphics[width=8.5cm,angle=0]{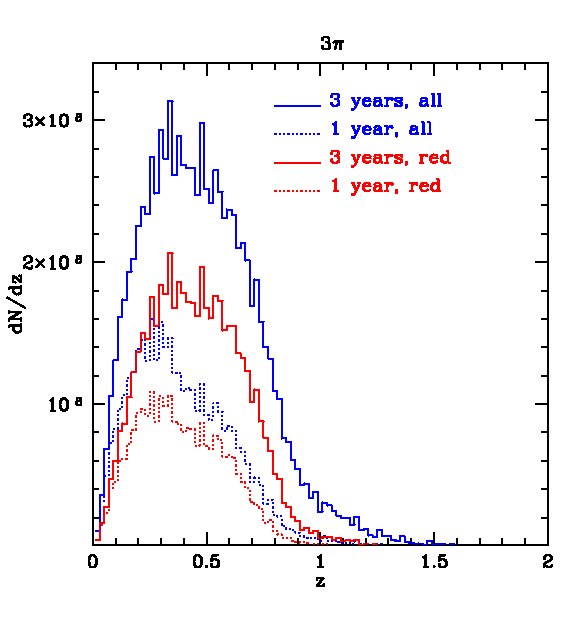}
\includegraphics[width=8.5cm,angle=0]{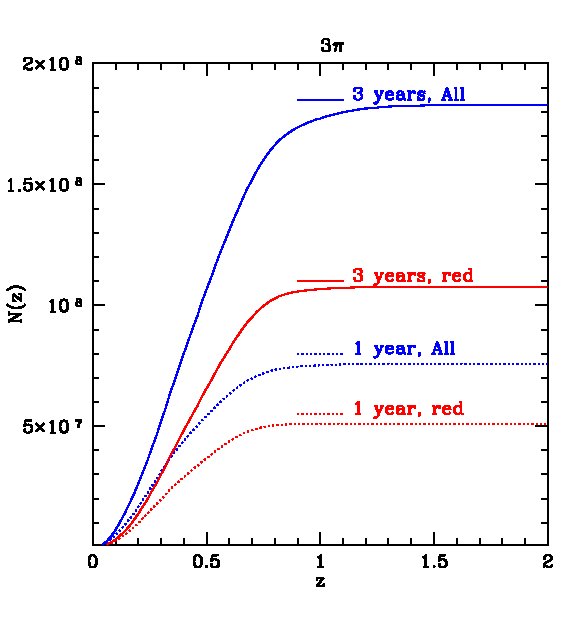}

\includegraphics[width=8.5cm,angle=0]{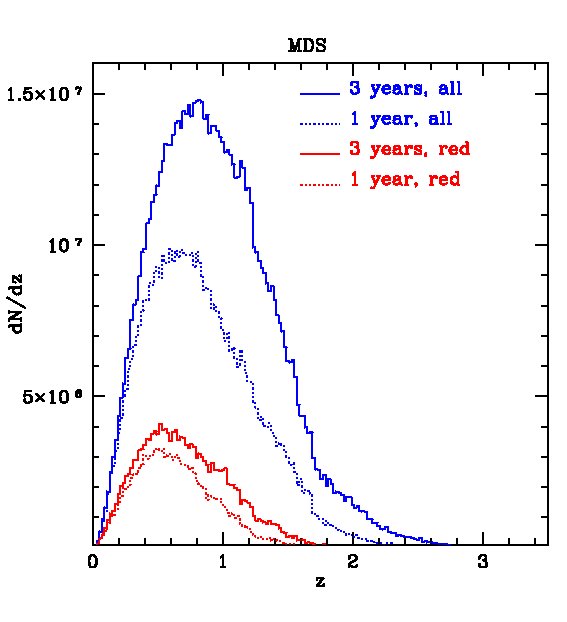}
\includegraphics[width=8.5cm,angle=0]{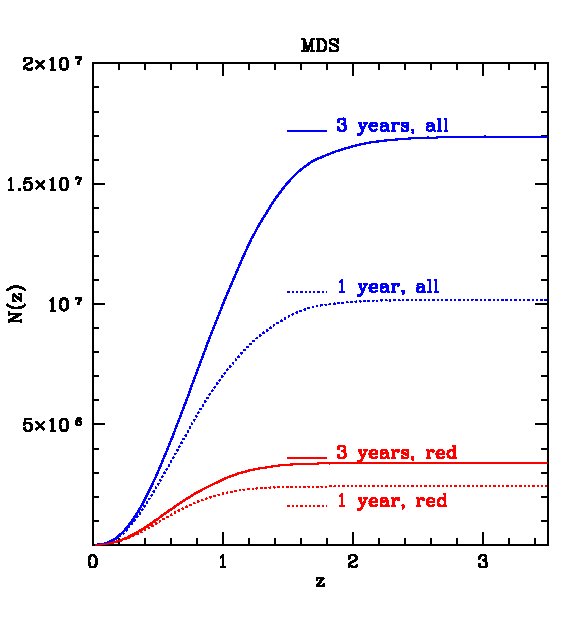}
\caption{Expected galaxy redshift distributions for galaxies detected 
in all 5 ($g,r,i,z,y$) PS1 bands in the $3\pi$ survey (top panels) and 
the Medium Deep Survey (MDS)
(bottom panels), as predicted by the {\sc galform} model. The
left-hand panels give the differential counts, in bins of $\Delta
z=0.02$. The right-hand panels give the cumulative counts. Blue lines
show results for all galaxies while the red lines refer exclusively to
red galaxies. Solid lines are for the 3-year surveys and dotted lines
for the 1-year surveys. Red galaxies are selected by a rest-frame
colour cut of $M_g-M_r>-0.04M_r-z/15-0.25$, where $z$ is the redshift
(see Fig.\ref{fig4}). Note that these predictions have been extrapolated 
from a mock catalogue which covers 10 square degrees, and so are 
noisier than would be expected for the actual survey sizes. 
\label{fig3}}
\end{center}
\end{figure*}

\begin{figure*}
\begin{center}
\includegraphics[width=8.5cm,angle=0]{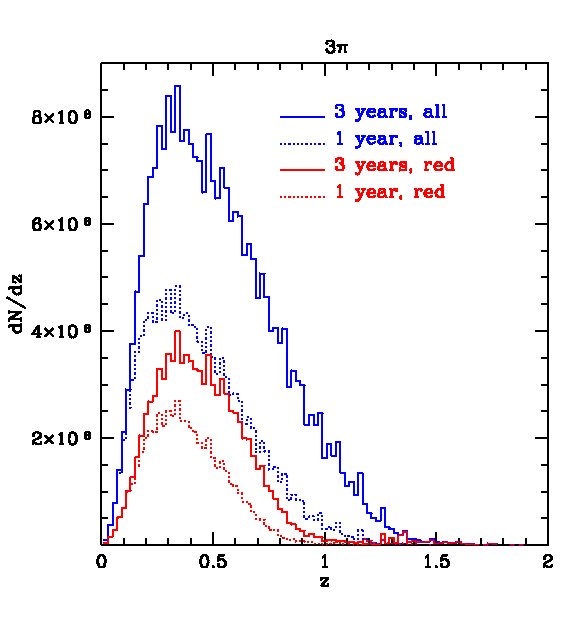}
\includegraphics[width=8.5cm,angle=0]{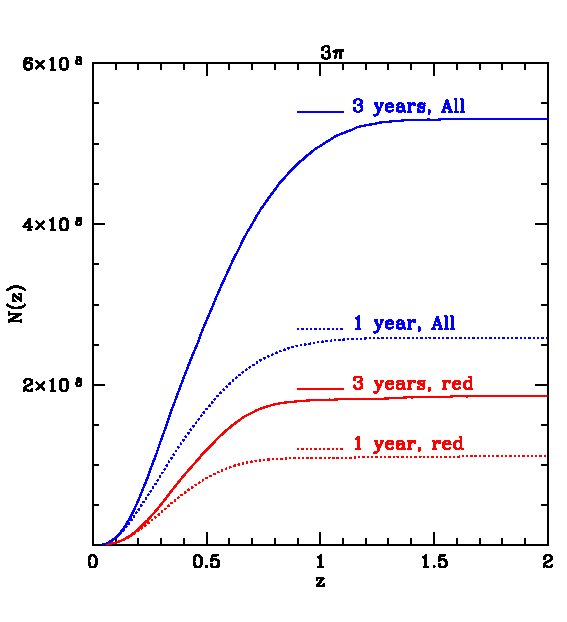}

\includegraphics[width=8.5cm,angle=0]{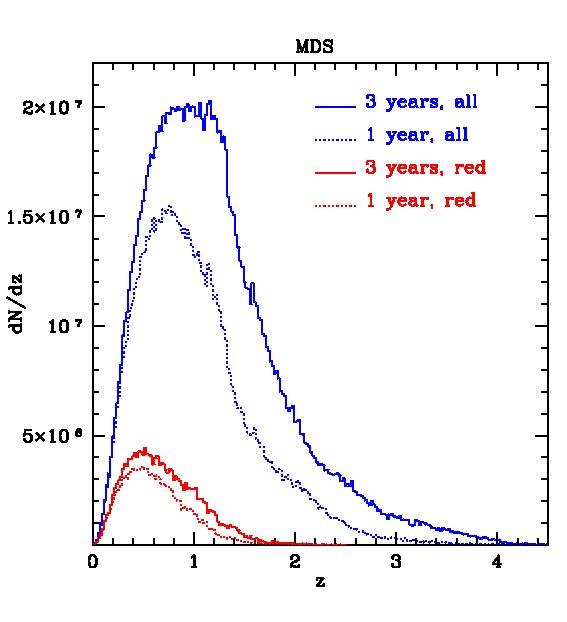}
\includegraphics[width=8.5cm,angle=0]{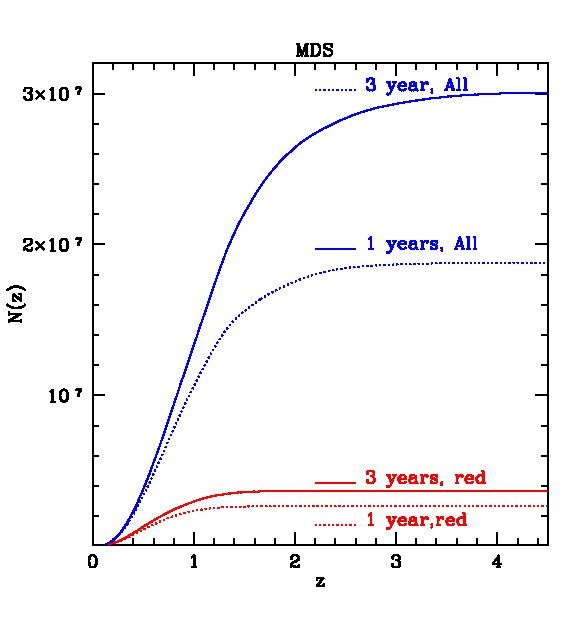}
\caption{As Fig.\ref{fig3} but for galaxies 
required to be detected only in the $g, r, i$, and $z$ bands. Without
requiring the shallow
$y$-band detection the number of galaxies is about twice as large
as with the full $g, r, i, z$ and $y$ constraints.
\label{fig3.1}}
\end{center}
\end{figure*}

\begin{figure*}
\begin{center}
\includegraphics[width=8.5cm,angle=0]{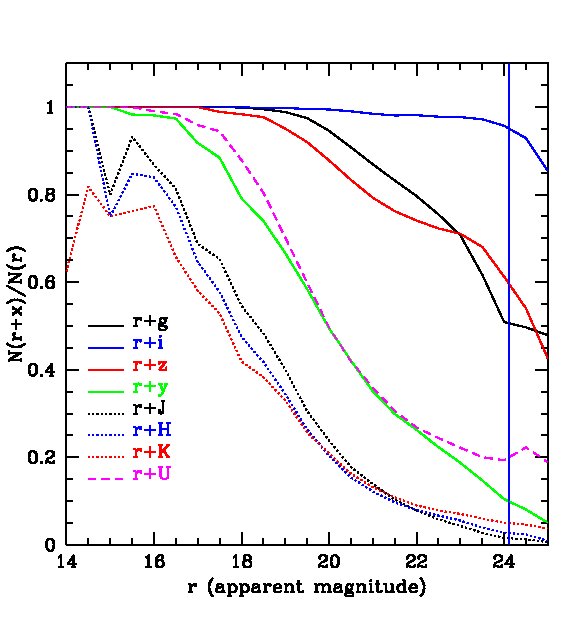}
\includegraphics[width=8.5cm,angle=0]{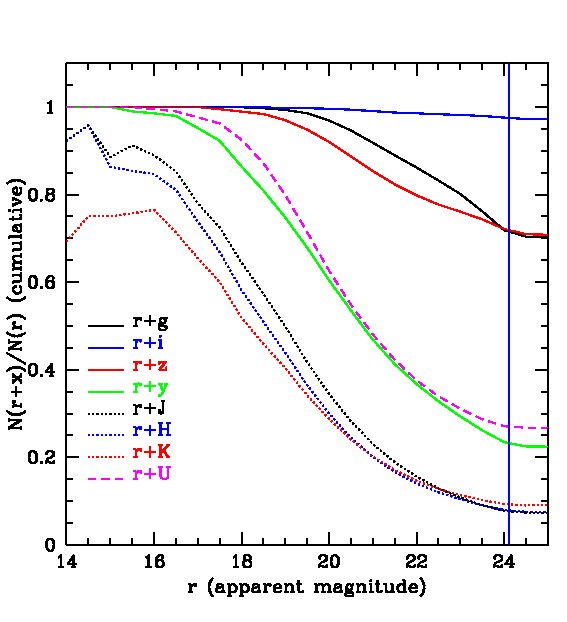}
\caption{Ratio of differential (left) and cumulative (right) counts for 
galaxies selected using a combination of
$r$-band and one other filter to the counts of galaxies selected using
the $r$-band alone, as a function of $r$-band magnitude. For the
additional filters, we use the UKIDSS $J, H$ and $K$ bands, the PS1
 $g, i, z$, and $y$ bands and a $U$-band . For the PS1 $grizy$
system, we adopt the third year magnitude cuts and for the UKIDSS
bands the LAS limits; for the $U$-band we assume a limit of
23~mag. The label $r+x$ denotes that galaxies are selected by
combining the $r$-band and one of the other bands. The vertical 
blue lines indicate the $r$-band 5$\sigma$ point source
detection limit after the 3-year surveys.
\label{fig6}}
\end{center}
\end{figure*}

\subsection{The magnitude limits for the  PS1 $3\pi$ and MDS surveys}

The signal registered on a CCD chip from a point source with total
apparent magnitude $m$, after an exposure time of $t$ seconds is:
\begin{equation}\label{eq0}
S=0.5~t\times 10^{-0.4(m-m_1)}, 
\end{equation}
where $m_1$ is the magnitude that produces $1$ electron per
second. The factor of $0.5$ comes from
assuming the PSF is a 2D Gaussian profile and integrating over the
FWHM of this profile.

The signal-to-noise ratio for a point source is given by:
\begin{equation}\label{eq1}
S/N=S/\sqrt{\sigma^2_{\rm P}+\sigma^2_{\rm S}+\sigma^2_{\rm RN}+\sigma^2_{\rm D}},
\end{equation}
where $\sigma^2_{\rm P}=0.5t \times10^{-0.4(m-m_1)}$ is the Poisson counting
noise for a source of magnitude $m$ observed for $t$ seconds;
$\sigma^2_{\rm S}=\frac{\pi}{4}\omega^2\times10^{-0.4(\mu-m_1)}t$ is the
variance from the sky background, where $\mu$ is the average sky
brightness in magnitudes per square arcsec and $\omega$, assumed to be
$0.78$ arcsecs, is the FWHM of
the PSF; $\sigma^2_{\rm RN}=\frac{\pi}{4}\omega^2\times A^2\times N_{\rm
read}^2$ is the read-out noise of the detector, where, for PS1,
A=3.846 pixels/arcsec and $N_{\rm read}=5$ is the read-out noise in
electrons; $\sigma^2_{\rm D}$ is the variance due to dark current and will
be assumed to be zero
\citep{Chambers06}. Table~\ref{tab1} lists the
expected values of the parameters $\mu$ and $m_1$
and also
gives the $5\sigma$ point source magnitude limits
resulting from applying this formula to the 3$\pi$ survey and the MDS
after one and three years.

The signal-to-noise for resolved extended sources will be smaller.
To estimate this we take
the Petrosian radius and the redshift of a galaxy and obtain the
solid angle subtended by $2r_P$ of the galaxy, $\theta_g$.
Then for extended sources, $\theta_{\rm g}> \omega$,
we define the signal and the noise to be the values integrated
over the source aperture $\theta_g$ rather than the FWHM of the
PSF. Thus the signal is simply
$S=t\times 10^{-0.4(m-m_1)}$, the Poisson noise $\sigma^2_{\rm P}=t
\times10^{-0.4(m-m_1)}$, the sky background variance is
$\sigma^2_{\rm S}=\frac{\pi}{4} \theta_{\rm g}^2\times10^{-0.4(\mu-m_1)}t$
and the read-out noise is
$\sigma^2_{\rm RN}=\frac{\pi}{4}\theta_{\rm g}^2\times
A^2\times N_{\rm read}^2$.  Since we have not convolved the image with
the PSF, this treatment would produce a sharp transition in the
noise level at the PSF limit. This can be avoided by approximating the
convolved diameter of the image by $\left(\theta_g^2+\theta_P^2
\right)^{1/2}$ and using this to replace $\theta_g$ in the
expressions
for $\sigma^2_{\rm S}$ and $\sigma^2_{\rm RN}$.
Table~\ref{tab1} gives the $5\sigma$ point source magnitude limits
resulting from applying this formula to the 3$\pi$ survey and the MDS
after one and three years.

\subsection{A test of the PS1 mock catalogues}

Before discussing predictions from our mock catalogues for the $3\pi$
and MDS PS1 surveys, we first carry out a simple test of the realism
of our mock catalogues. The {\sc galform} semi-analytic model has been
shown to be consistent with various basic properties of the local
galaxy population such as the luminosity functions in the $b_J$ and
$K$ bands
\citep{Cole01,Norberg02,Huang03}.
The B06 version of the model also gives an excellent match to the
evolution of the rest-frame $K$-band luminosity function, including
the data from the K20 \citep{Pozzetti03} and {\sc munic} surveys
\citep{Drory03} up to redshift $z=1.5$
\citep{Bower06}. 

Since neither the $b_J$ nor the $K$-band coincide with any of the PS1
$grizy$ bands, for a more direct test we compare predicted galaxy
number counts in the $r$-band with data. At the faint end we use the
number counts over 5 sq deg from the {\sc deep}2 survey
\citep{Coil04}, which are complete to 24.75 in
the $R$ band. To minimise sample variance at the bright end, we 
use galaxy number counts in the SDSS commissioning data
\citep{Yasuda01} which cover about 440~sq deg and
are complete to $r^*=21$. (We have checked that the difference
between the commissioning data and more recent SDSS releases 
\citep{Fukugita04, Yasuda07} is negligible). We
compute the {\sc galform} model predictions, including uncertainties,
from 10 realizations of 10 sq deg mock surveys. The results, displayed
in Fig. \ref{fig2}, show that our model prediction agrees very well
with both the {\sc deep}2 and the SDSS datasets. Note that we have
applied the 0.15 magnitude shift discussed in \S2.2 to the model
galaxies.

\subsection{Expected PS1 galaxy numbers counts and redshift distributions}

We now discuss the expected population statistics for the PS1 surveys
predicted by our mock catalogues. We apply Petrosian magnitude cuts in
each of the PS1 bands and plot the expected galaxy number counts in
0.5~magnitude bins in Fig.~\ref{fig5}, for both the 3-year $3\pi$
survey and the MDS. The figure shows that, with the Petrosian
magnitude cuts, the samples are no longer complete to the 5$\sigma$
point source magnitude limits in the various bands, but rather only to
$\sim$2 magnitudes brighter. Note that the $y$-band magnitude limit is
substantially shallower than the others and so, if one requires
detection in all five bands, the $y$ limit is the most restrictive.

The cumulative distributions on the right hand panels of
Fig.~\ref{fig5} reveal the staggering number of galaxies that will be
detected by PS1. For example, in the $g$-band after 3 years, we expect
about $10^9$ galaxies in the 3$\pi$ survey and nearly $10^8$ in the
MDS.

The expected redshift distributions for the two surveys are shown in
Fig.~\ref{fig3} for galaxies detected in all 5 bands and in
Fig.~\ref{fig3.1} for galaxies detected only in $g, r, i$ and $z$,
i.e. not requiring the shallow $y$-band detection.
In the first case, the $n(z)$
distribution peaks at $z\sim 0.5$ for the $3\pi$ survey, with about
$8\times10^7$ and $1.8\times10^8$ galaxies detected (in all 5 bands)
in the 1- and 3-year surveys respectively. The survey is so huge,
that, after 3 years, we expect about 10 million galaxies at $z>0.9$
and 5 million at $z>1$. 
For the MDS
survey, the $n(z)$ distribution peaks at $z\sim 0.8$, with a total of
$1.7\times 10^7$ galaxies after 3 years of which around 0.5~million
lie at $z>2$. Removing the $y$-band constraint leads to a large
increase in the number of galaxies, as shown in Fig.~\ref{fig3.1}. In
this case, the $3\pi$ survey will contain $\sim 5\times10^8$ galaxies
after three years, with about 30 million at $1<z<1.3$, while the MDS
will contain $\sim 3\times10^7$ galaxies, with 4 million at $z>2$.

For certain applications, for example, for the estimate of photometric
redshifts discussed in the next section, it might be desirable to
supplement the PS1 $grizy$ filter system with other bands,
particularly in the near infrared. The UKIDSS Infrared Deep Sky Survey
\citep[e.g.][]{Lawrence07,Hewett06} is
particularly relevant in this context.  The UKIDSS Large Area Survey
(LAS) aims to map about 4000~sq deg of the Northern sky (contained
within the $3\pi$ survey) over the course of a few hundred nights. The
Deep Extragalactic Survey (DXS) aims to cover 35~square degrees of the
sky in three separate regions which have a large overlap with the
fields chosen for the MDS. Details of the $J$, $H$ and $K$ magnitude
limits of the UKIDSS surveys are listed in Table~\ref{tab2}. In order
to assess the compatibility of the PS1 and UKIDSS surveys, we show in
Fig.~\ref{fig6} the reduction in galaxy counts, relative to a pure
$r$-band selection, that would result from combining in turn each of
the filters with the $r$-band filter. We see, once again, that the
$y$-band cut (green line) is much shallower than the other PS1
bands. The UKIDSS (LAS) $J, H$ and $K$ bands are even
shallower. Combining $r$-band and $U$-band detections also results in
a large reduction in the counts even for an optimistic $U$-band limit
of 23 mag. Fig.~\ref{fig6} suggests that, in spite of the large area
overlap with the $3\pi$ survey, the UKIDSS (LAS) survey may be too
shallow to pick up PS1 galaxies at high redshifts. It will be very
difficult for a $U$-band survey to pick up a significant number of PS1
galaxies.

\begin{figure*}
\begin{center}
\includegraphics[width=8.5cm,angle=0]{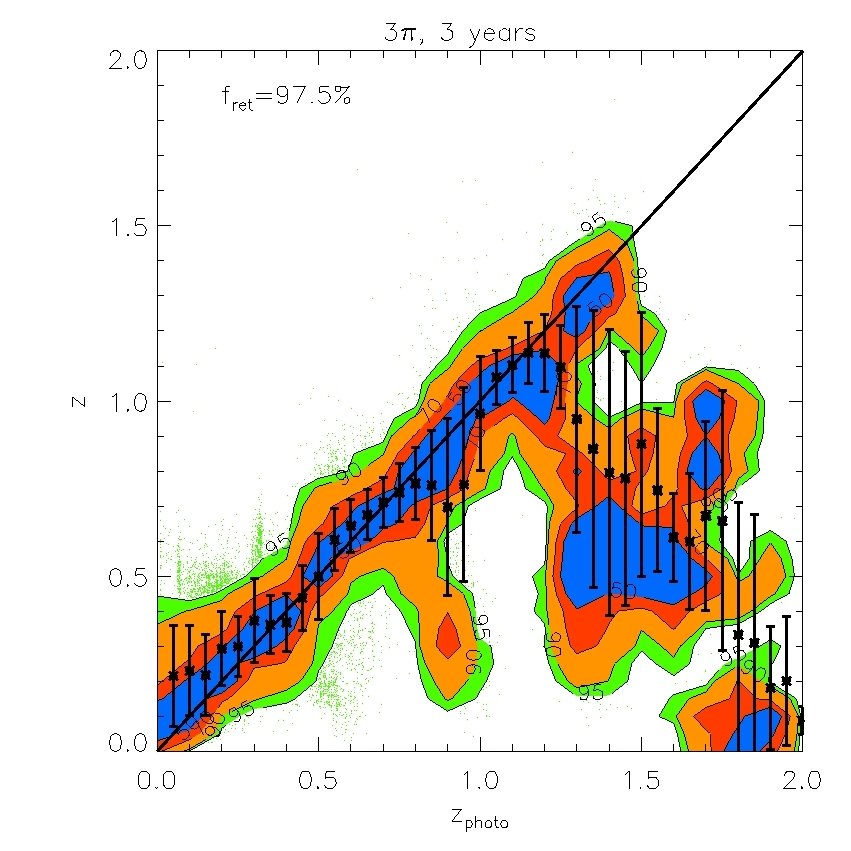}
\includegraphics[width=8.5cm,angle=0]{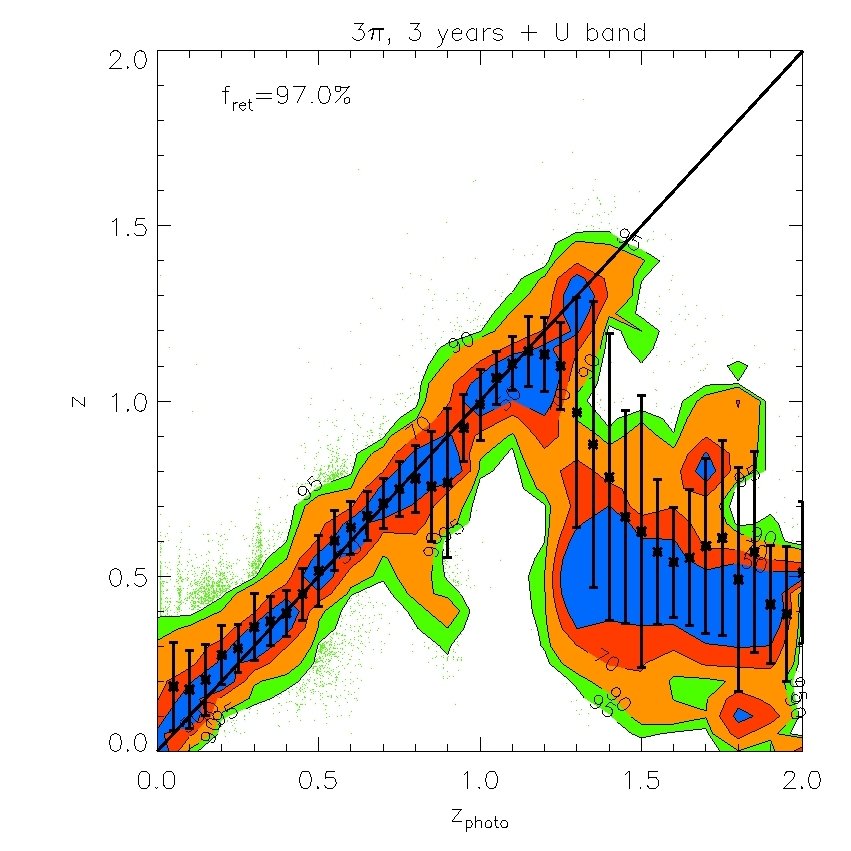}

\includegraphics[width=8.5cm,angle=0]{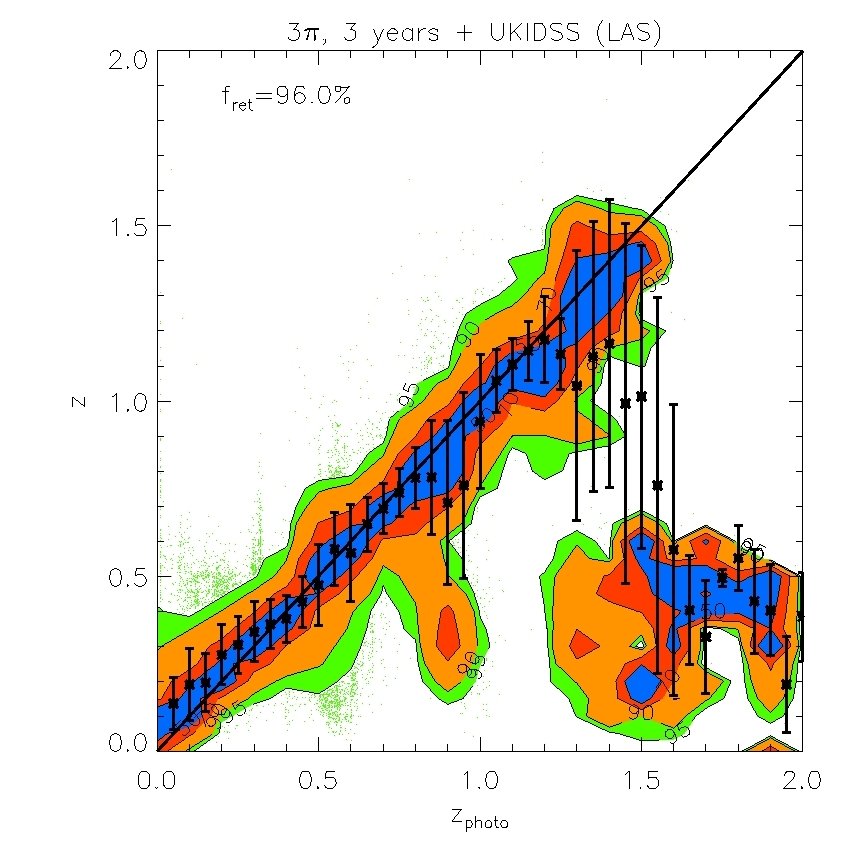}
\includegraphics[width=8.5cm,angle=0]{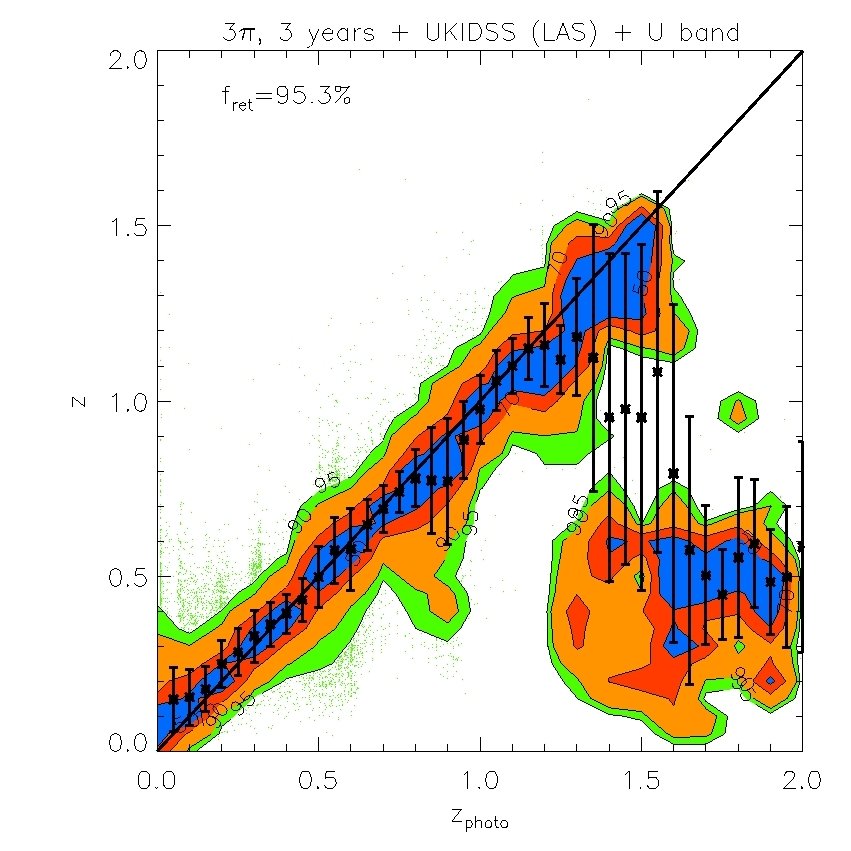}
\caption{True (``spectroscopic'') redshifts plotted against photometric
redshifts in a 10 sq deg mock PS1 $3\pi$ 3-year galaxy
catalogue. In each bin of photo-$z$ the contours show the regions
containing 50\% (blue), 70\% (red), 90\% (orange) and 95\% (green) of
the galaxies. Galaxies with true redshifts falling outside the 
$95\%$ contours are shown by the green dots. 
Galaxies are selected by applying 5$\sigma$ Petrosian
magnitude cuts for all 5 PS1 $grizy$ bands. If the flux in some other
filters ($U$, $B$, $J$, $H$ or $K$) drops below its $5\sigma$ limit, 
the detected flux is still used with its uncertainty. The error bars 
show the {\it rms} scatter
after $3\sigma$ clipping. The percentages of galaxies retained after
the clipping are given in the legend. Top left: PS1 $grizy$ band
data only. Top right: PS1 $grizy$ combined with $U$-band. Bottom
left: PS1 $grizy$ combined with UKIDSS (LAS) $J$ and $K$. Bottom
right: PS1 $grizy$, $U$-band and UKIDSS (LAS) $J$ and $K$.
\label{fig7}}
\end{center}
\end{figure*}
\begin{figure*}
\begin{center}
\includegraphics[width=8.5cm,angle=0]{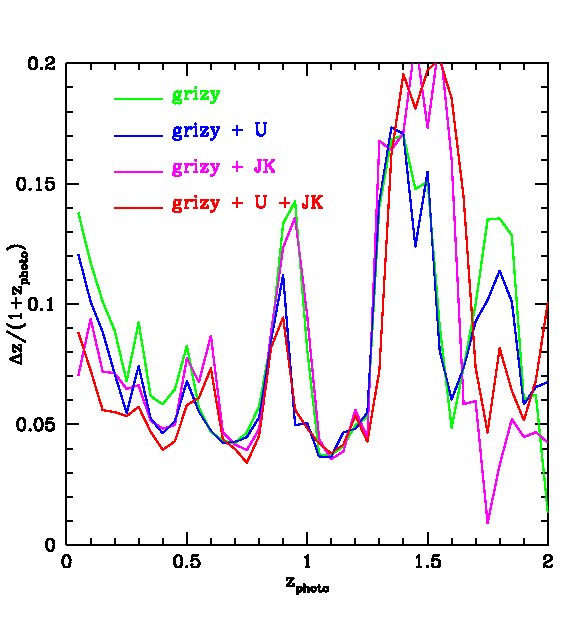}
\includegraphics[width=8.5cm,angle=0]{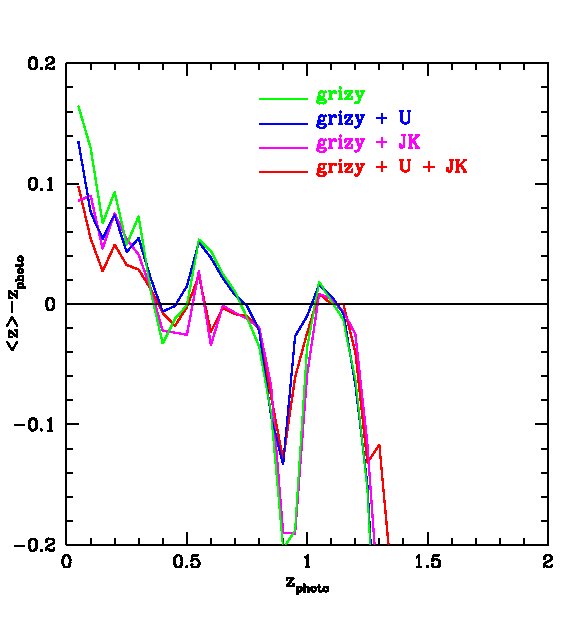}
\caption{Accuracy of the photometric redshift estimates in a 10 sq deg
mock PS1 $3\pi$ 3-year galaxy catalogue. Only galaxies remaining after
applying a 3$\sigma$ clipping procedure to the binned data are
retained in the estimate.
The retained fractions are given in the legend of
Fig.~\ref{fig7}.
We use a
redshift bin size of $\Delta z=0.05$. Left panel: $1\sigma$
uncertainty divided by ($1+z$) plotted against the photo-$z$ 
Right panel: Systematic deviation of the mean photometric redshift in
each bin from the true value, as a function of photo-$z$. 
\label{fig9}}
\end{center}
\end{figure*}

\begin{figure*}
\begin{center}
\includegraphics[width=8.5cm,angle=0]{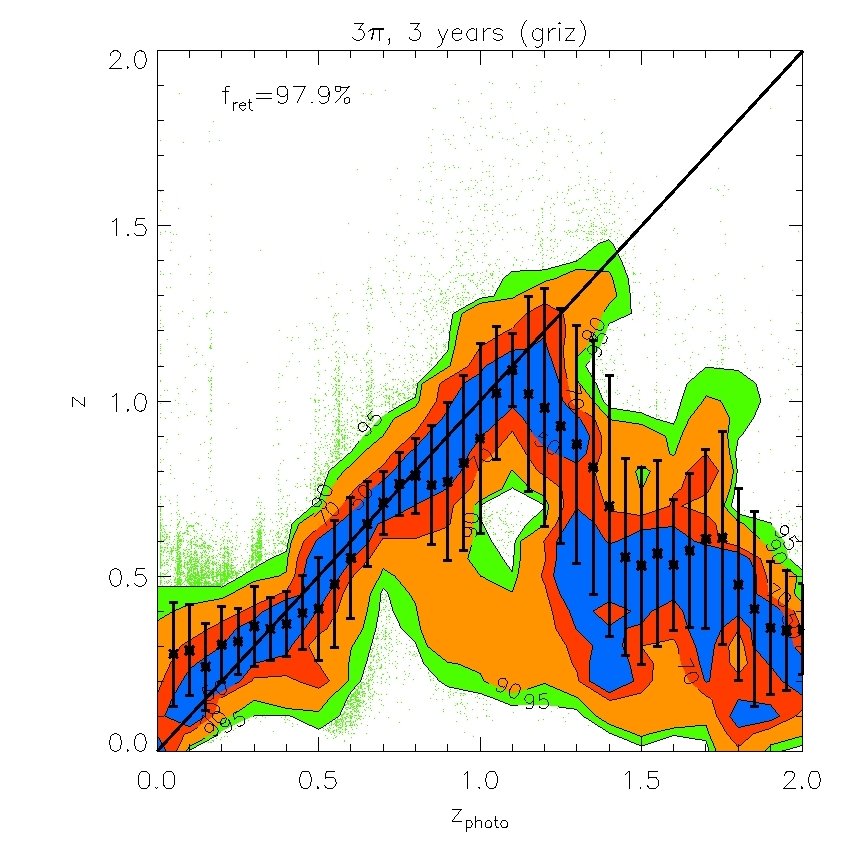}
\includegraphics[width=8.5cm,angle=0]{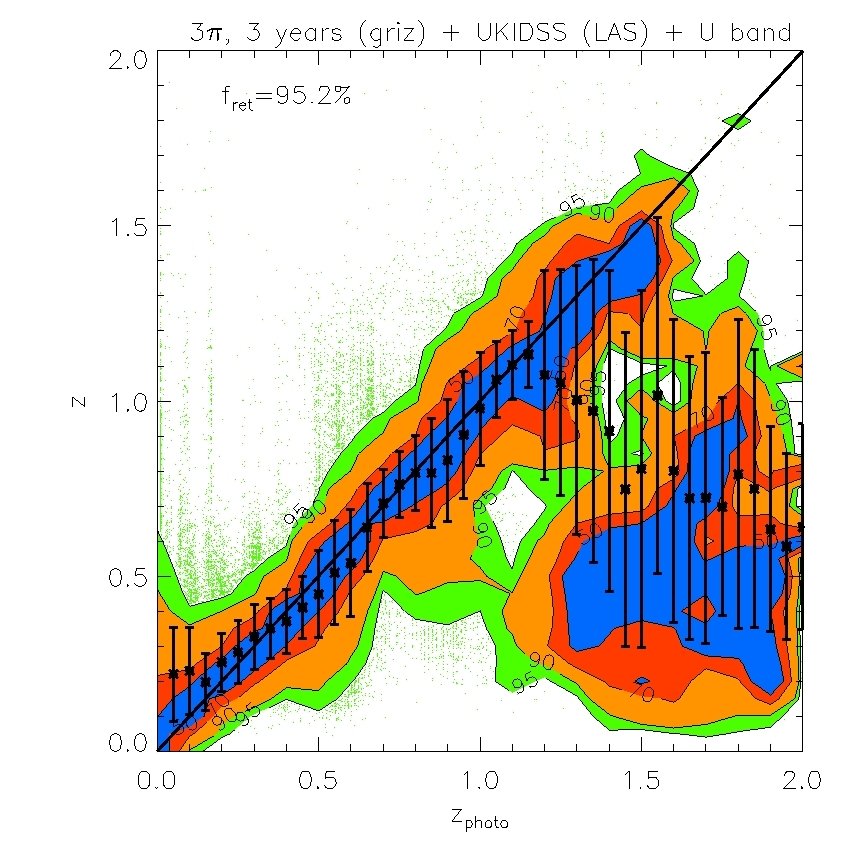}
\caption{As Fig.~\ref{fig7}, but using a larger, deeper sample by not
requiring a $y$ band detection and
using only $griz$ fluxes in the
determination of photometric redshifts. Without the $y$-band, more galaxies
are detected, but the error and bias in the photometric redshifts increase.
Left panel: results when using only the PS1 photometry.
Right panel: results when adding UKIDSS (LAS)
$J, K$-band and with $U$-band photometry.
\label{fig7.0}}
\end{center}
\end{figure*}

\begin{figure*}
\begin{center}
\includegraphics[width=8.5cm,angle=0]{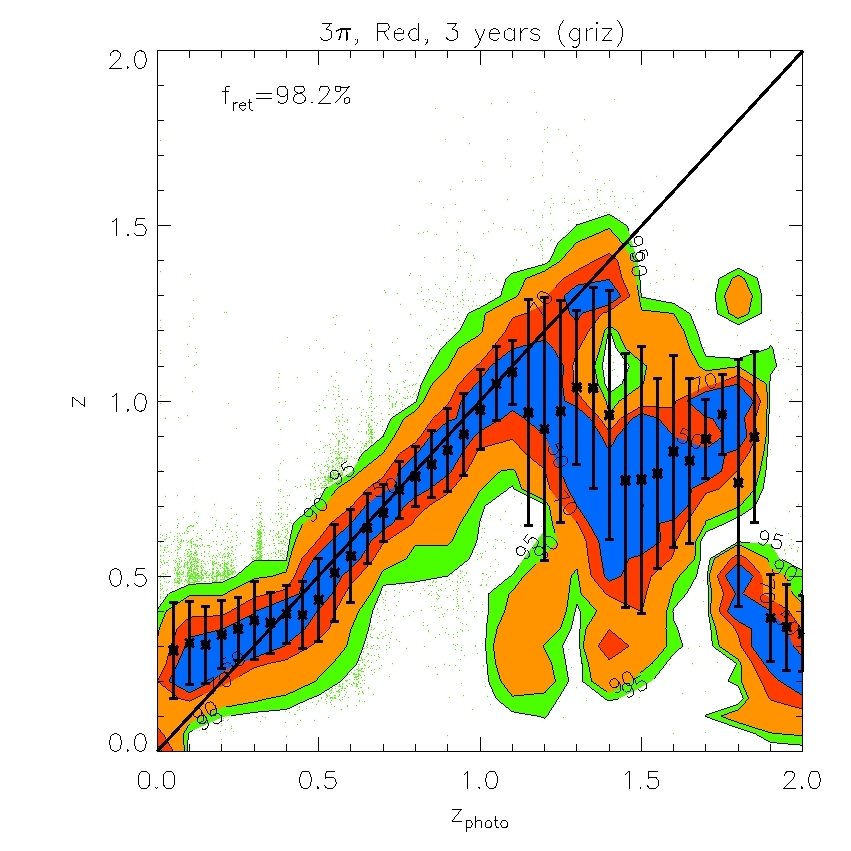}
\includegraphics[width=8.5cm,angle=0]{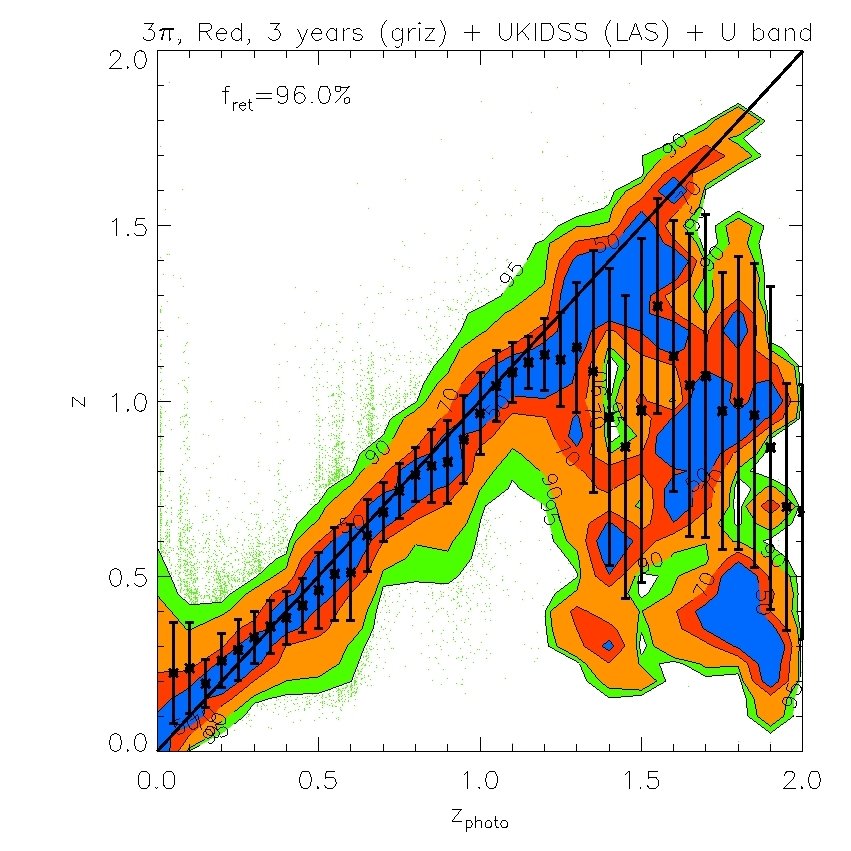}
\includegraphics[width=8.5cm,angle=0]{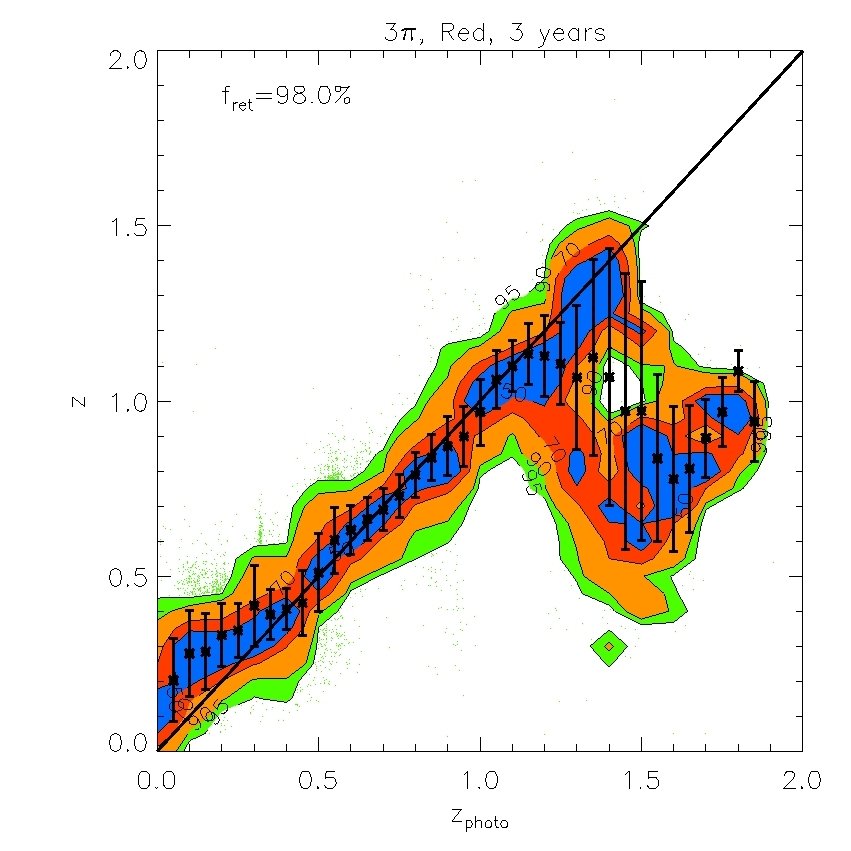}
\includegraphics[width=8.5cm,angle=0]{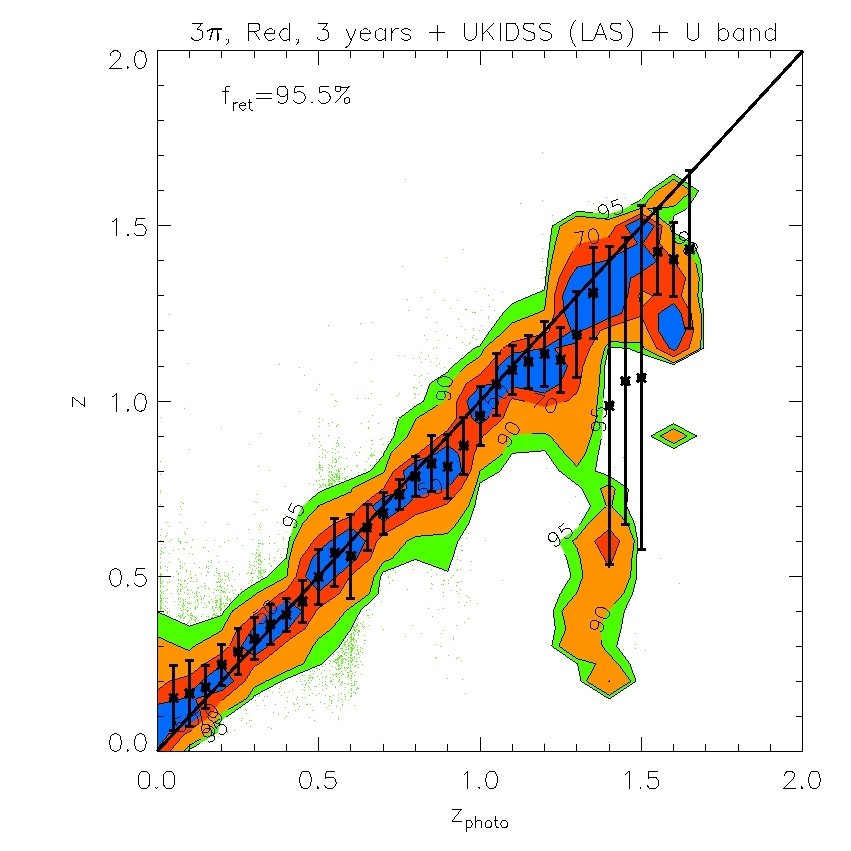}
\caption{``Spectroscopic'' versus photometric redshifts, as in
Fig.~\ref{fig7} and Fig.~\ref{fig7.0}, but for galaxies that are
required to be red in their rest frame $g-r$ colour.
Top panels: deep samples in which no $y$-band detection is
required. The left hand panel makes use of only PS1 $griz$ photometry,
while the right hand panel makes use of additional
UKIDSS (LAS) $J, K$-band and fiducial $U$-band photometry.
Bottom panels: These panels show the results for the shallower
sample in which detections in all 5 $(grizy)$ PS1 bands are required.
Again the left hand panel uses only PS1 data and the right hand panel
makes use of additional
UKIDSS (LAS) $J, K$-band and fiducial $U$-band photometry.
\label{fig7.1}}
\end{center}
\end{figure*}

\begin{figure*}
\begin{center}
\includegraphics[width=8.5cm,angle=0]{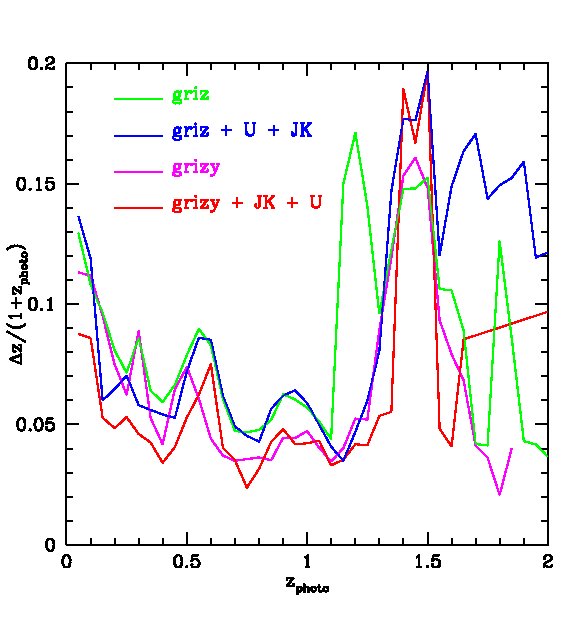}
\includegraphics[width=8.5cm,angle=0]{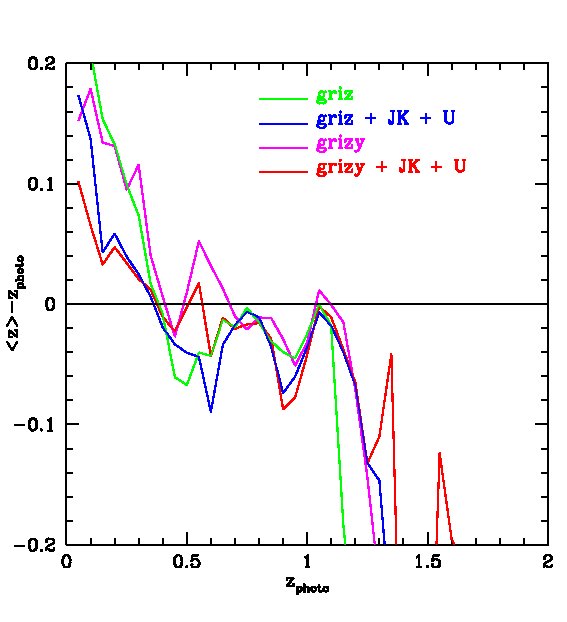}
\caption{
Accuracy of the photometric redshift estimates in a 10 sq deg
mock PS1 $3\pi$ 3-year red galaxy catalogue. Results using only 
$griz$ fluxes in the determination of photometric redshifts 
are shown together. Without the $y$-band, more galaxies detected, 
but the error and bias in the photometric redshifts increase.
Only galaxies remaining after applying a 3$\sigma$ clipping procedure 
to the binned data are plotted. We use a
redshift bin size of $\Delta z=0.05$. Left panel: $1\sigma$
uncertainty divided by ($1+z$) plotted against the photo-$z$. 
Right panel: Systematic deviation of the mean photometric redshift in
each bin from the true value, as a function of photo-$z$. 
\label{fig9.1}}
\end{center}
\end{figure*}

\begin{figure*}
\begin{center}
\includegraphics[width=8.5cm,angle=0]{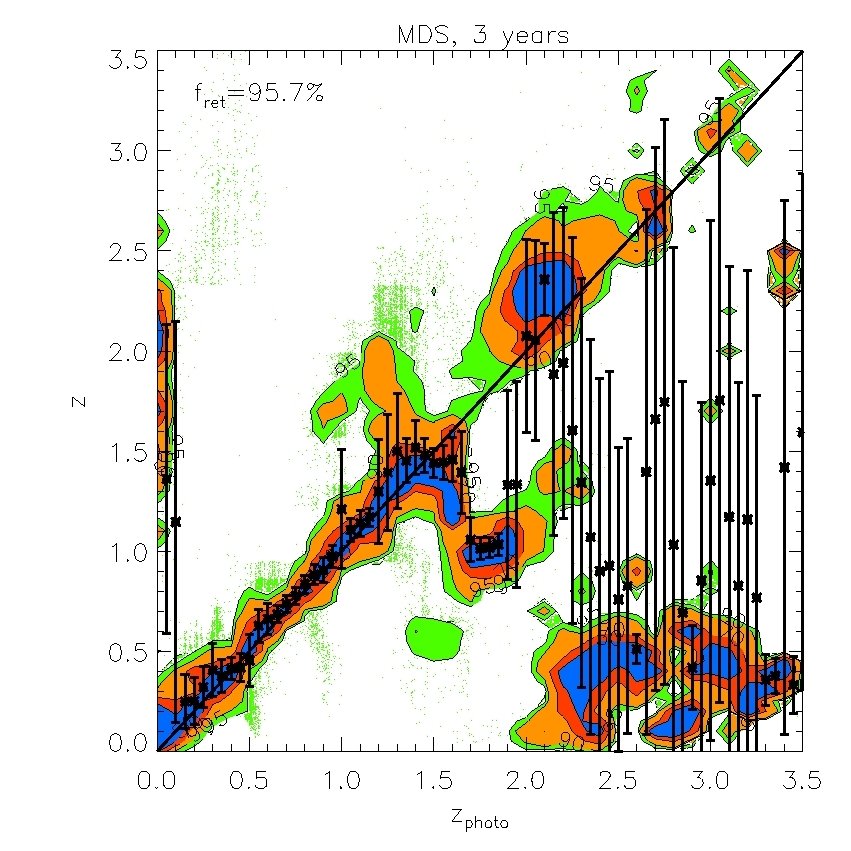}
\includegraphics[width=8.5cm,angle=0]{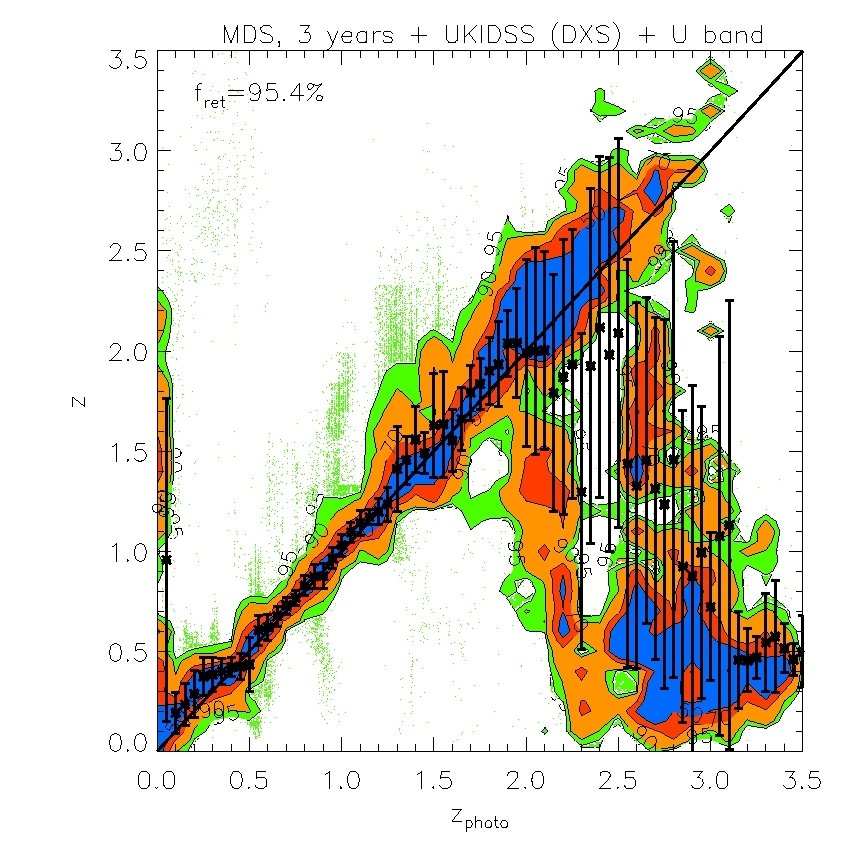}
\includegraphics[width=8.5cm,angle=0]{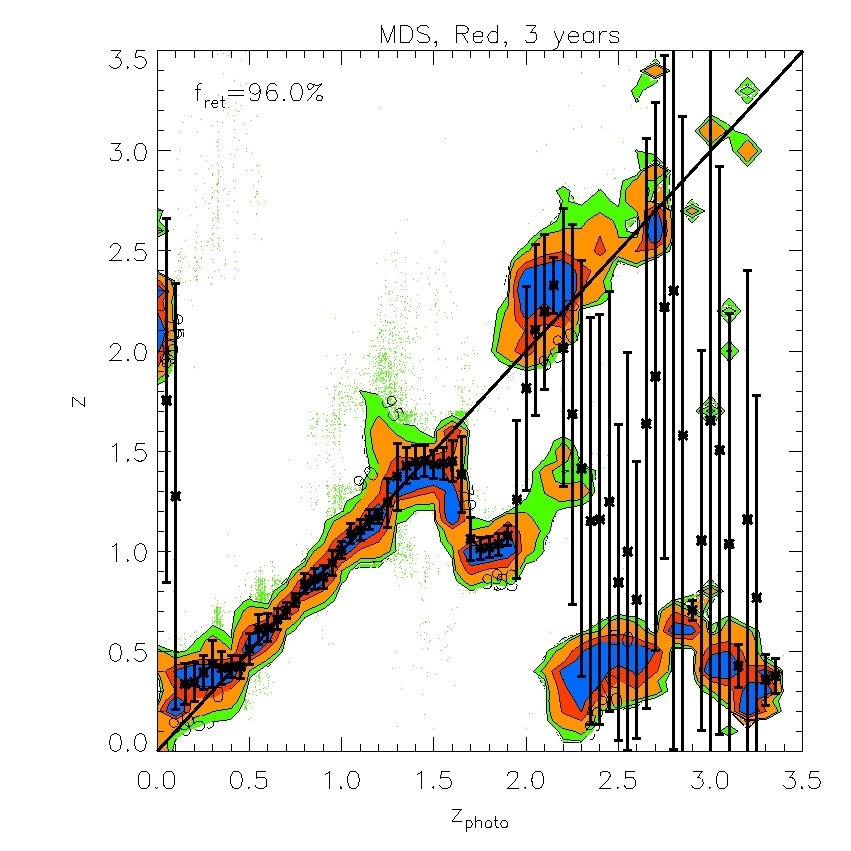}
\includegraphics[width=8.5cm,angle=0]{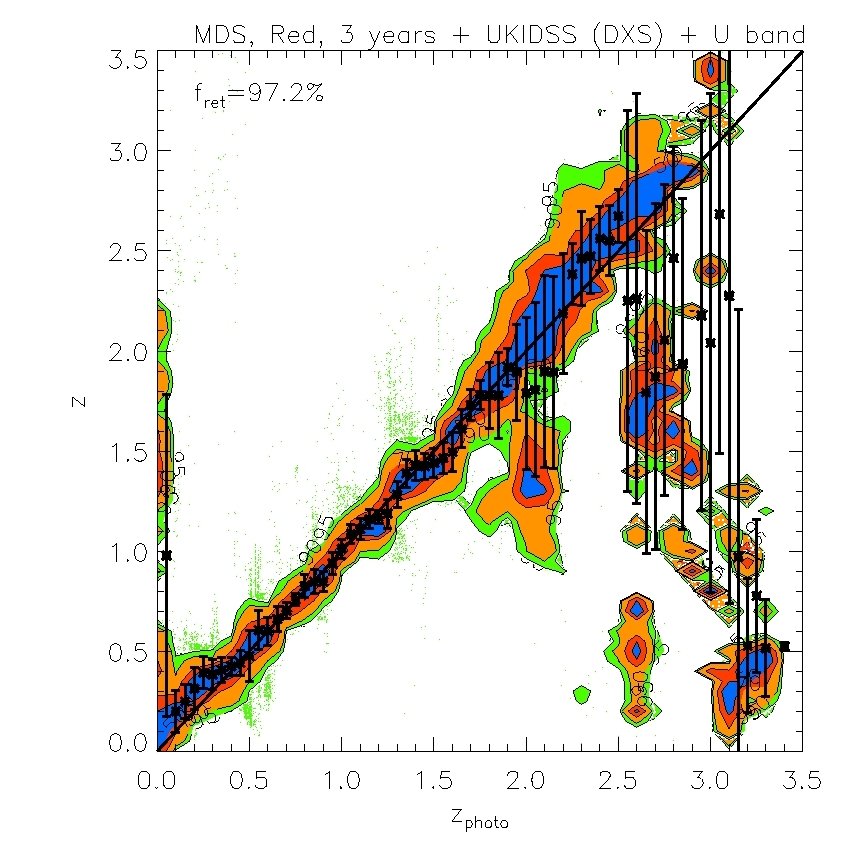}
\caption{True (``spectroscopic'') redshifts plotted against photometric 
redshifts for the 3-year MDS survey.  The data are presented in the
same fashion as in Figs~\ref{fig7},\ref{fig9} and~\ref{fig10}, 
but for the MDS we extend the redshift
range to $z=3.5$.
Top panels: predictions for the 3-year MDS using 1 sq deg mock
catalogues. Bottom panels: predictions for samples of red
galaxies. Left panels, results by using only the $grizy$ 
photometry. Right Panels, results by adding UKIDSS (DXS)
$J, K$-band and with $U$-band photometry. 
Galaxies are selected applying 5$\sigma$ Petrosian magnitude cuts for all 5 PS1
$grizy$ bands. If the flux in some other filters ($U$, $B$, $J$, $H$
or $K$) drops below its $5\sigma$ limit, the detected flux is still used with 
its uncertainty. The error bars show the {\it rms} scatter
after $3\sigma$ clipping. The percentages of galaxies retained after
the clipping are given in the legend.
\label{fig10}}
\end{center}
\end{figure*}
\begin{figure*}
\begin{center}
\includegraphics[width=8.5cm,angle=0]{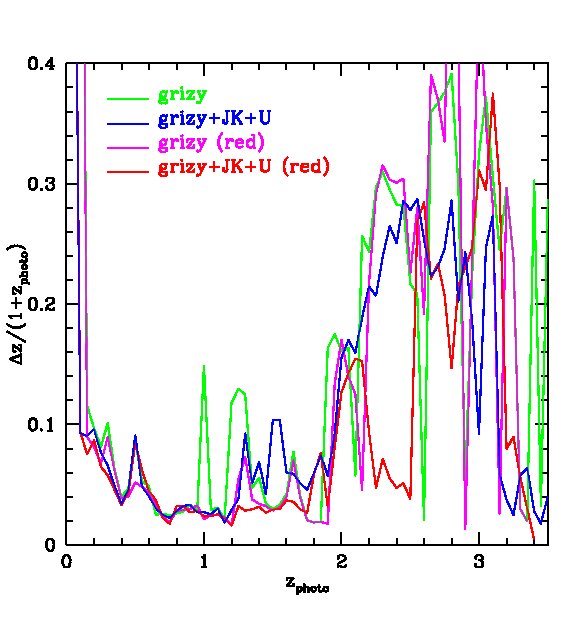}
\includegraphics[width=8.5cm,angle=0]{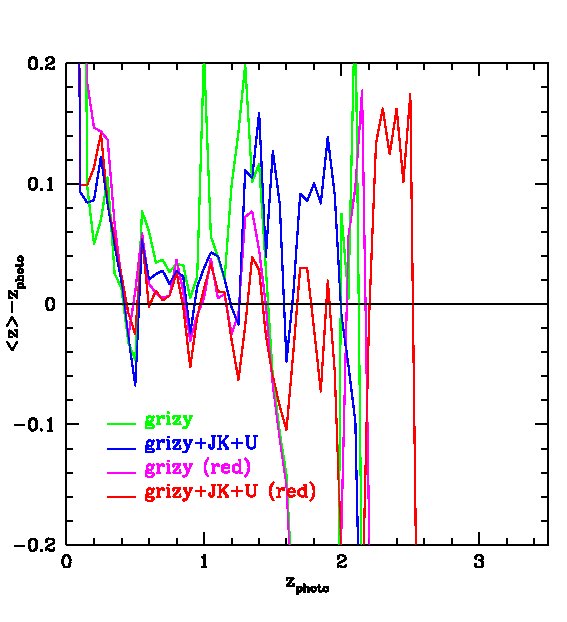}
\caption{Accuracy of the photometric redshift estimates for the 3-year MDS 
survey shown in Fig.~\ref{fig10}. Only galaxies remaining after
applying a 3$\sigma$ clipping procedure to the binned data are
plotted. Galaxies are binned in the spectroscopic redshift axis with bin 
size $\Delta z=0.05$. Left panel: $1\sigma$ uncertainty divided by 
($1+z$) plotted against the photo-$z$.
Right panel: Systematic deviation of the mean photometric redshift in 
each bin from the true value, as a function photometric redshift. 
\label{fig11}}
\end{center}
\end{figure*}

\section{Photometric redshifts in the PS1 survey}

We now examine the accuracy with which redshifts are likely to be
estimated using PS1 photometry. For this purpose we adopt an 
off-the-shelf photometric redshift code, the Hyper-$z$ code of 
\citet{Bozonella00} which is based on fitting template spectra. 
We do not attempt to tune the performance of the estimator in anyway, 
and so our results should perhaps be regarded as providing a 
pessimistic view of the photometric redshift performance of PS1.   
Once PS1 data become available, bespoke estimators will be developed 
which are optimized to return the smallest random and systematic errors 
for the PS1 filter set and galaxies by having empirically adaptive
galaxy templates \citep[e.g.][]{Bender01}. 

The basic principle behind the template fitting approach to photometric 
redshift estimation is the following. The observed SED of a 
galaxy is compared to a set of template spectra and a standard
$\chi^2$ minimisation is used to obtain the best fit:
\begin{equation}
\chi^2(z)=\sum^N_{i=1}\left[\frac{F_{\rm{obs},i}-b\times
F_{\rm{tem},i}(z)}{\sigma_i}\right], 
\end{equation}
where $F_{\rm{obs},i}$, $F_{\rm{tem},i}$ and $\sigma_i$ are the observed fluxes,
template fluxes and the uncertainty in the flux through filter $i$,
respectively and $b$ is a normalization factor. For the fitting
procedure, we input the PS1 $grizy$-band filter transmission curves
and, when appropriate, those of the UKIDSS near infrared bands and a
$U$ band filter.  We consider different reddening laws and two
sets of model templates: the mean
spectra of local galaxies given by
\citet[CWW]{Coleman80} and the synthetic spectra given by
\citet[BC]{Bruzual93}. We set a redshift range  
of $0<z<3$ for the 3$\pi$ and $0<z<4$ for the MDS sample.

One might be concerned that the use of Bruzual \& Charlot stellar population 
synthesis models to generate both the template spectra and the galaxy spectra 
in the mock catalogues might lead to an underestimate of the error on the 
photometric redshift. There are two key differences between the mock spectra 
and the templates which mean that this is not an issue: i) the complexity of 
the composite stellar populations of mock galaxies and ii) the differing  
treatments of dust extinction. The template spectra correspond to a single 
parameter star formation history (characterized by an exponentially decaying 
star formation rate, where the $e$-folding time is treated as a parameter) 
and a fixed metallicity for the stars. The mock galaxies, on the other hand, 
have complicated star formation histories which cannot be fitted by a 
decaying exponential (see Baugh 2006 for examples of star formation histories 
predicted by the semi-analytical models). Furthermore, the stars in the mock 
galaxy have a range of metallicities. Hyper-$z$, in common with many other 
photometric redshift estimators, assumes that dust forms a foreground 
screen in front of the stars with a particular extinction law. In {\sc galform}, 
the dust and stars are mixed together. This more realistic geometry can lead 
to dust attenuation curves which look quite different from those assumed in 
the photometric redshift code \citep{Granato00}.

The Hyper-$z$ code calculates a redshift probability distribution, $P(z)$, for
each galaxy. Because of a degeneracy between the 4000~${\rm{\AA}}$ and the 
912~$\rm{\AA}$
breaks, the shape of $P(z)$ can have a double peak, causing some low redshift
galaxies to be misidentified as high redshift galaxies and viceversa. Some of
these misidentifications can be removed by applying extra constraints, for
example, the galaxy luminosity function and the differential comoving volume as
a function of redshift \citep{Mobasher07}.
For a given
observed flux, both these functions provide an estimate of the probability that
the galaxy has redshift $z$ which can be use to modulate $P(z)$. The highest
peak in the combined probability distribution gives the best estimate of the
photometric redshift. We use the $r$-band luminosity function of the B06 model
for this purpose.

We now discuss how the accuracy and reliability of the photometric
redshift estimates depends on various choices. We do this by
calculating photometric redshifts for a 10 sq deg subsample of our
mock PS1 $3\pi$ 3-year catalogue and comparing these with the true
redshifts (which we will sometimes refer to as the ``spectroscopic''
redshifts.)

\noindent {\it 1. Choice of SED template (CWW {\it vs} BC)}

Our tests show that using 5 input spectral types: burst, S0, Sa, Sc
and Im, gives good results; adding more spectral types does not
produce further significant improvement.  We find that fitting with
the CWW templates gives larger statistical uncertainties and systematic
deviations from the true redshift than fitting with the BC templates,
especially at high redshift ($z>1$).  The reason for this could be
that the CWW templates are based on observations of the local universe
and may not be sufficiently representative of galaxies at high
redshift. 
In what follows, we will exclusively use the BC templates.

We also experimented with BC templates for different
metallicities. Because of the age-metallicity degeneracy in galaxy
SEDs, we did not find any improvement by allowing the metallicity to
vary while letting the age of the stellar populations be a free
parameter. Since the 4000~$\rm{\AA}$ break only becomes detectable after a
stellar population has aged beyond $10^7$ years, we exclude templates
with ages smaller than this. This greatly improves the results for low
redshift galaxies ($z<0.5$).

\noindent {\it 2. Dependence on photometric bands}

The accuracy of the redshift estimates depends on the choice of
photometric bands. With our 10 sq deg 3-year mock catalogues, we can 
explore which combination of bands gives optimal results for PS1.
We have considered many combination of the PS1 $grizy$ photometry
with UKIDSS (LAS) $JHK$ and fiducial $B$ and $U$ photometry.
Note that,
if the flux through any of the $U$, $B$,
$J$, $H$ or $K$ filters drops below the 5$\sigma$ flux limit, the 
noisy measured flux is still used with its appropriate uncertainty.

We find that the $B$-band (whose effective wavelength is very close to
the $g$ band) does not improve the fits if $U$ is available, but the
$U$-band is still useful even when $B$-band data are included. We also
find that the $H$-band is not important provided $J$ and $K$ are
available. However, both $J$ and $K$ are important for improving the
quality of the fits. Therefore, in what follows we will ignore $B$ and
$H$. Our results are displayed in Figs.~\ref{fig7} and~\ref{fig9}. In
Fig.~\ref{fig7}, we plot the ``spectroscopic'' redshifts  against
our estimated photometric redshifts for the 4 cases above. 
For clarity, rather than plotting each galaxy on these plots, we have
instead displayed contours that indicate the region in each bin of
photo-$z$ that contains 50\%, 70\%, 90\% and 95\% of the galaxies.
Galaxies with ``spectroscopic'' redshifts 
falling outside the $95\%$ contours are shown individually by green dots. 
To evaluate the $1\sigma$ scatter we eliminate extreme outliers through standard
3$\sigma$ clipping. Typically over $f_{\rm{ret}}=95\%$ of the galaxies
are retained, as indicated in the legend.  Fig.~\ref{fig7} plots the
true or ``spectroscopic'' redshift against our estimated photometric
redshift for the PS1 $grizy$ photometry alone and when supplemented by
$U$-band photometry, UKIDSS (LAS) photometry, or both.
Fig.~\ref{fig9}
(left panel) shows $\Delta z/(1+z)$
plotted against redshift where $\Delta z$ is the $1\sigma$ error from
Fig.~\ref{fig7}. The bias in the mean of each redshift bin relative to
the true value is also shown (right panel).

The PS1 $grizy$ bands alone give relatively accurate photometric
redshifts in the range $0.25<z<0.8$, with typical {\it rms} values of $\Delta
z/(1+z)\sim 0.06$. The random and systematic errors increase at both lower and
higher redshifts and there is a population of low redshift ($z<1$) galaxies
which are incorrectly assigned high redshifts. Adding the $U$-band
produces only a moderate improvement at all redshifts. Using both the $J$ and
$K$ bands  results in a significant improvement at $z<0.5$, but not at
higher redshifts. Finally, combining the $U$, $J$ and $K$, produces the
best results. For this best case, the {\it rms} error, $\Delta z/(1+z)\sim
0.05$, in the range $0.5<z<1$ and, for $z<1.2$, it is never larger than 0.15.
\begin{figure}
\resizebox{\hsize}{!}{
\includegraphics{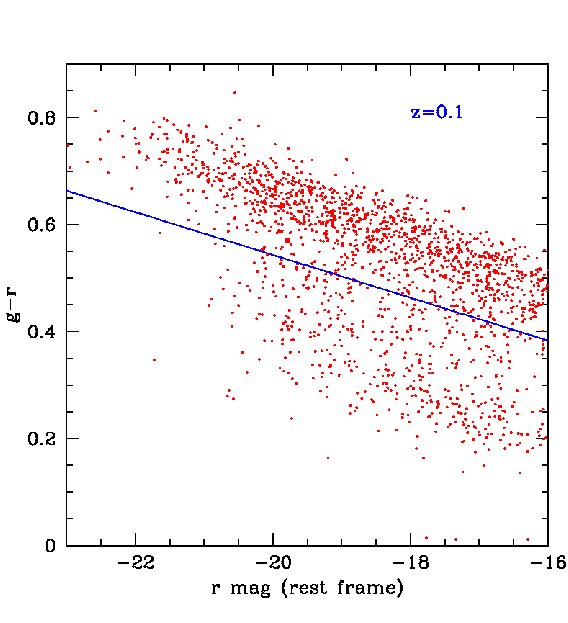}
\includegraphics{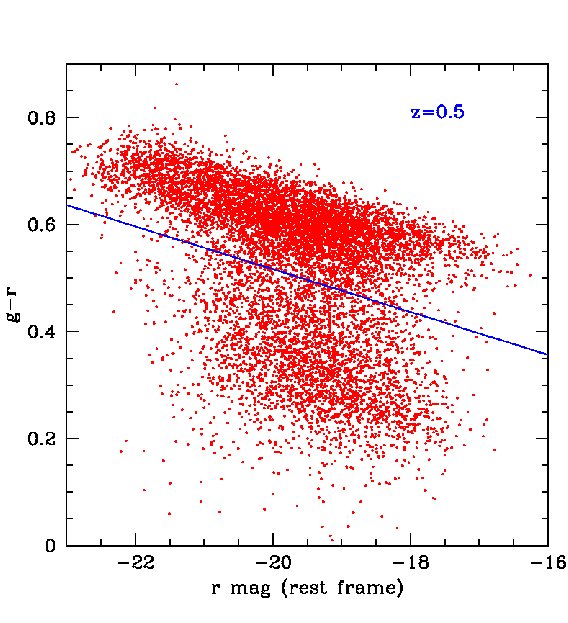}}
\resizebox{\hsize}{!}{ 
\includegraphics{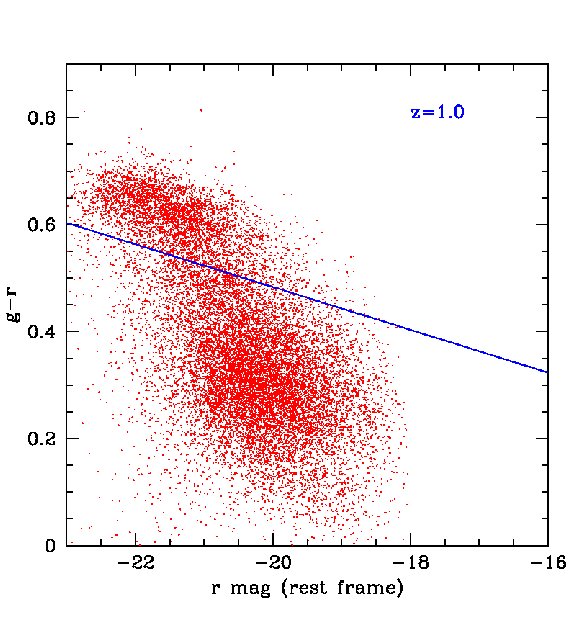}
\includegraphics{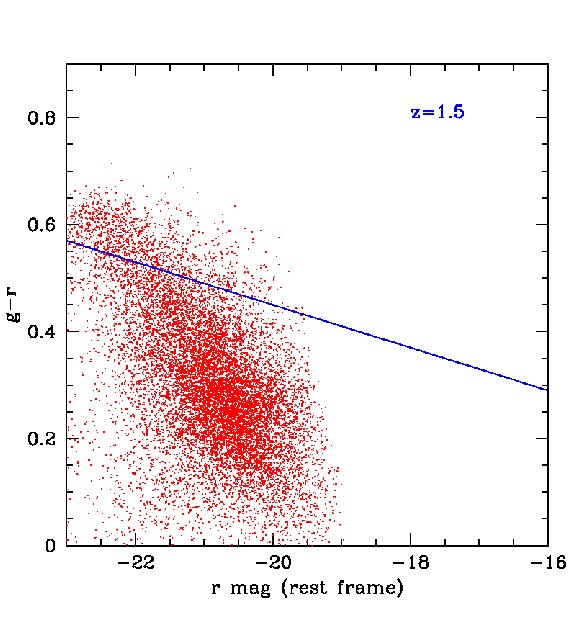}}
\caption{Expected colour-magnitude relation for the MDS 3-year mock
catalogue. The plots show  rest-frame $g-r$ colour {\it versus} rest-frame $r$-band
magnitude predicted by {\sc galform} at the redshifts given in each panel. The
blue line is $M_g-M_r=-0.04M_r-z/15.0-0.25$, where $z$ is the redshift. 
Galaxies above the line make up the ``red'' sample.
\label{fig4}}
\end{figure}

\begin{figure}
\includegraphics[width=8.5cm,angle=0]{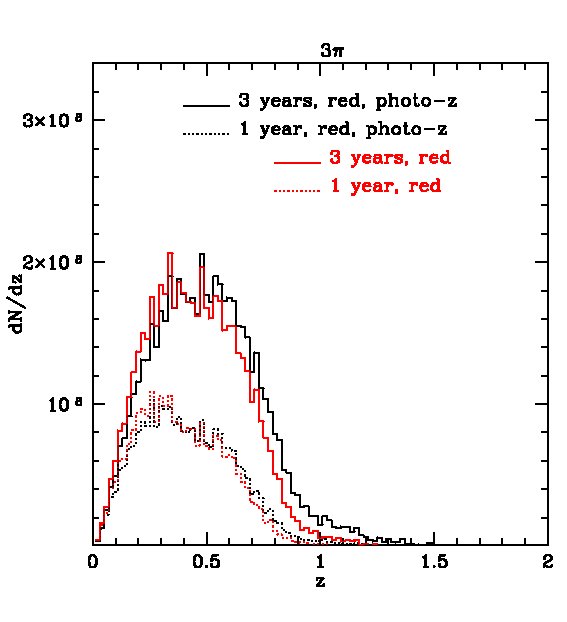}
\caption{Predicted redshift distributions for ``red'' galaxies in the PS1 
$3\pi$ survey, selected in two different ways. The red lines show results for a
sample selected by rest-frame $g-r$ colour (according to
$M_g-M_r>-0.04M_r-z/15-0.25$); the black lines show results for a sample
selected by the best fit photo-$z$ spectral type, with detail in 
the text. The redshift bin is $\Delta z=0.02$. The good agreement
between the two selection methods suggests that it may be possible to 
select the red sample directly from the observed photometry.
\label{fig7.2}}
\end{figure}

We saw in \S3.3 that requiring that galaxies be detected in $y$, the
shallowest PS1 filter, reduces the sample size by factors of
2-3. The deeper sample that we achieve by only requiring $griz$
detections has significantly less accurate photo-$z$s. This is
shown in Fig.~\ref{fig7.0}, in which 
we measure photometric redshifts using only $griz$ photometry.
In this, the {\it rms} in the redshift range $0.25<z<0.8$ increases 
from $0.06$ to $0.075$ and the bias changes little. 

Photometric redshift estimates for the MDS are shown in the top 2
panels of Fig.~\ref{fig10} and their accuracy is quantified by the
green and blue lines in Fig.~\ref{fig11}. If only the PS1 
$grizy$ are available, an accuracy of $\Delta z/(1+z)\sim 0.05$ is 
achievable for $0.02<z<1.5$. Adding the UKIDSS (DXS) and the $U$-band 
improves the estimates considerably, but there is still a clear bias 
at very low and high redshifts. This is mainly because the depths of the
UKIDSS (DXS) and our assumed $U$ band is insufficient to match the
depth of the MDS so faint galaxies are not detected in the UKIDSS $J$
and $K$ band nor in the $U$ band.

For certain applications, for example, the measurement of baryonic
acoustic oscillations discussed in the next section, smaller {\it rms}
errors than those found above are required. These can be achieved by
selecting subsamples of galaxies whose spectra are particularly well
suited for the determination of photometric redshifts, such as red
galaxies which have strong 4000~$\rm\AA$ breaks. The most direct way to
define a red subsample is by using the rest frame $g-r$ colours. In
Fig~\ref{fig4}, we plot the predicted rest-frame $g-r$ against
$r$-band luminosity at four different redshifts in our mock MDS
catalogue. A cut at $M_g-M_r>-0.04M_r-z/15.0-0.25$ neatly separates
out the red sequence, particularly at $z<1$. The redshift
distributions of red galaxies defined this way are shown by the red
lines for both the 3$\pi$ survey and the MDS in Figs.~\ref{fig3}
and~\ref{fig3.1}. The distributions peak at slightly lower redshifts
than the full samples, but there is still an impressive number of red
galaxies in the two surveys. For example, in the 3$\pi$ survey we
expect about 200 million galaxies after 3-years if detection in $y$ is
not required or 100 million if it is.

In practice, rest-frame $g-r$ colours are difficult to estimate from
the observations. An alternative method for identifying red galaxies
is to use the spectral type determined by Hyper-$z$. We define a red 
sample by the following criteria, galaxies which are best fit with 
the `Burst' spectral type and a stellar population older than $10^9$yr.

Fig.~\ref{fig7.2} shows the redshift distribution for this
sample which can be seen to be very similar to the redshift
distribution of a red sample selected by rest-frame $g-r$ colour. This
suggests that it will be possible to select a red galaxy sample
directly from the observational data alone.

Fig.~\ref{fig7.1} and Fig.~\ref{fig10} show 
photometric redshift estimates for red galaxies in the 3-year 3$\pi$ 
survey and the MDS respectively. Their accuracy is illustrated by 
the magenta ($grizy$ photometry only) and red lines ($grizy$+$JK$+$U$) in 
Figs.~\ref{fig9.1} and Fig.~\ref{fig11} respectively. Results without the 
$y$-band photometry are shown in the top panels of Fig.~\ref{fig7.1} and in
green ($griz$ photometry only) and blue ($griz$+$JK$+$U$) lines in 
Fig.~\ref{fig9.1}. These figures show the dramatic improvement in
photometric redshift accuracy for red galaxy samples. For example, for
the $3\pi$ survey, the {\it rms} value of $\Delta z/(1+z)$ can be as
low as 0.02 at $z \sim 0.8$ when combining $grizy$ with UKIDSS (LAS) 
and $U$ bands measurements. Similarly, in the MDS with the same combination of 
filters, the accuracy for red galaxies is much higher than for the 
sample as a whole and can be as good as $\Delta z/(1+z) \sim 0.03$ in 
the redshift range $0.75<z<2.5$. 

\begin{figure*}
\begin{center}
\includegraphics[width=8.5cm,angle=0]{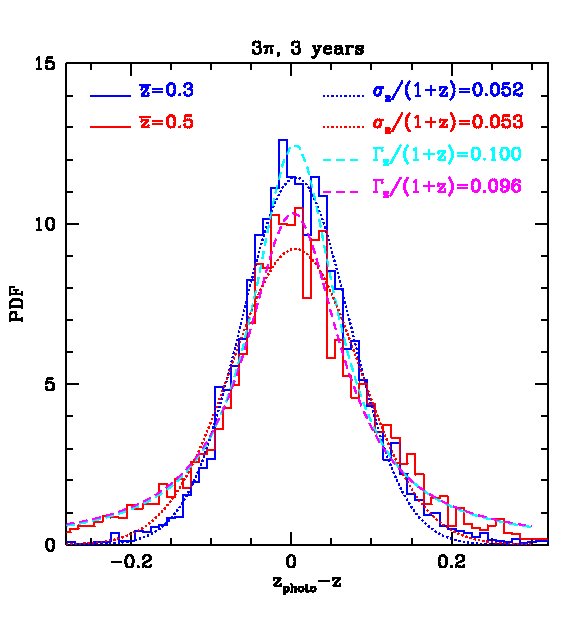}
\includegraphics[width=8.5cm,angle=0]{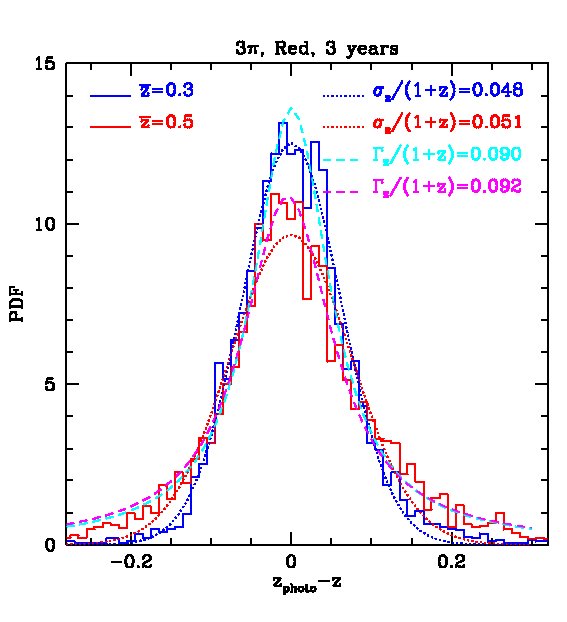}
\caption{
The distribution of photo-$z$ errors at redshift $z\sim 0.3$ and 
$z \sim 0.5$ for the 3-year $3\pi$ galaxy catalogues. The histograms are 
normalized to integrate to unity. 
Histograms in blue ($z\sim 0.3$) and red ($z\sim 0.5$) show the errors 
resulting from combining the $grizy$ bands 
with UKIDSS (DXS) $J, K$-band and with $U$-band photometry. 
They could be equally well fitted by Gaussian and Lorentzian 
distributions. $\sigma_z$ is the $rms$ width of the Gaussian function 
and $\Gamma_z$ is the FWHM of the Lorentzian function. Dotted lines 
show the best-fit Gaussians and the dashed lines illustrate the 
best-fit Lorentzian functions. Left: All galaxies, Right: Red galaxies.
\label{PhotozErr}}
\end{center}
\end{figure*}
Finally, we consider the form of the distribution of the photo-$z$ errors in 
Fig.~\ref{PhotozErr}. The photo-$z$ error distributions are well fitted by 
a Gaussian function, with variance $\sigma_z \approx \Delta_z$. 
The error distribution could also be equally well fitted by a 
Lorentzian function. Example distributions are shown at $z \sim 0.3$ and 
$\ \sim 0.5$ in Fig.~\ref{PhotozErr}. 
An application of our results for the size and form of the photo-$z$ 
errors is presented in the next section, 
in which we investigate their effect on the baryonic 
acoustic oscillation measurements.

\section{Implications for BAO Detection}

In this section we investigate the impact of using photometric
redshifts on the accuracy with which the baryonic acoustic oscillation
(BAO) scale can be measured from the power spectrum of galaxy
clustering. BAOs have been proposed as a standard ruler with which the
properties of the dark energy may be measured
\citep{Blake03,Linder03}. 
Our aim here is to provide a simple quantification of the factor by
which the effective volume of a survey is reduced when photometric
redshifts are used in place of spectroscopic redshifts. This will 
provide a rule of thumb indicator of the relative performance of 
photometric and spectroscopic surveys for the measurement of BAO. 
We defer a more extensive treatment of the full impact of the survey 
window function on the measurement of BAOs to a later paper. Mocks with 
clustering will play an important role in assessing the optimal way to 
measure the clustering signal in photometric surveys. 

The photometric redshift technique allows large solid angles of sky to
be covered to depths exceeding those accessible spectroscopically at a
low observational cost. However, the inaccurate determination of a
galaxy's redshift
results in an uncertainty in its position and this leads to a distortion
in the pattern of galaxy clustering. We shall refer to a measurement
of the power spectrum which uses photometric redshifts to assign
radial positions as being in ``photo-$z$" space.

The errors introduced by photometric redshifts can be modelled as
random perturbations to the radial positions of galaxies. 
As we have found from our photo-$z$ measurements that the photo-$z$ errors 
can be well fitted by a Gaussian function, if we assume that these 
perturbations are Gaussian
distributed
 with mean equal to the true redshift and width
$\sigma_{\rm z} \approx \Delta z$, then the Fourier transform of the measured
density field, $\delta_{\rm pz}(\underline{k})$, can be written
as
\begin{equation}
\delta_{\rm pz} (\underline{k}) = \delta_{\rm z} (\underline{k}) \,
\exp(-0.5 k_{\rm z}^2 \, \sigma_{\rm z}^2) ,
\end{equation}
where $k_{\rm z} = \underline{k} . \underline{\hat z} $,
$\underline{\hat z} $ is the line-of-sight direction
and $\delta_{\rm z}(\underline{k})$ is
the density field 
measured in redshift
space. From this expression, the spherically averaged power spectrum
can be approximately\footnote{It is an approximate expression since
the redshift space distortions and photometric redshift errors do not
commute under a spherical average \citep[see][]{Peacock94}.} 
written as:

\begin{equation}
P_{\rm pz} (k) = P_{\rm z} (k) \, \frac{\sqrt{\pi}}{2} \, \frac{\rm{Erf}(k
\, \sigma_{\rm z})}{ k \sigma_{\rm z}} , 
\label{pk_conv}
\end{equation}
\noindent where
${\rm Erf}(x) =\frac{2}{\sqrt{\pi}}\int_0^x \exp(-t^2) \rm{d}t$
is the error function.
In addition, the power spectrum in photo-$z$ space can be seen as that
in redshift space with additional damping on small scales due to the
large value of $\sigma_z$. On very large scales the main contribution
to the power spectrum comes from modes with wavelengths larger than
the typical size of the photometric redshift errors. Therefore, the
clustering on these scales is essentially unaffected. On the contrary,
on scales comparable to and smaller than the photo-$z$ errors,
structures are smeared out along the line-of-sight. The modes
describing these scales along the line-of-sight contain little
information about the true distribution of galaxies and contribute
only noise to the power spectrum.
\begin{figure*}
\begin{center}
\includegraphics[width=8.5cm,angle=0]{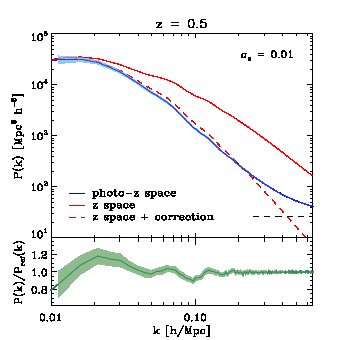}
\includegraphics[width=8.5cm,angle=0]{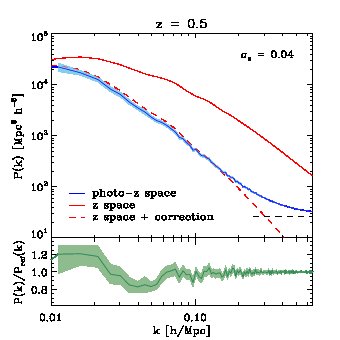}
\caption{The mean and standard deviation of the dark matter power spectrum
averaged over an ensemble of 50 N-Body simulations at $z=0.5$. The
top-panels display the power spectrum in three different cases: (i)
redshift space (solid red line), (ii) photo-$z$ space (blue line) in
which the position of each dark matter particle has been perturbed to
mimic the effect of photometric redshift errors, and (iii) the
photo-$z$ space power spectrum derived from Eq.~(\ref{pk_conv}) and
the measured redshift space power spectrum (red dashed lines).  The
horizontal dashed line illustrates the shot-noise level.  In the
bottom panels we plot the photo-$z$ power spectrum divided by a smooth
reference spectrum. This reveals the impact of photometric redshift
errors directly on the baryonic acoustic oscillations (BAO). An
increase in these errors causes an increase in the noise and a
decrease in the amplitude of the BAO at high wavenumber. This implies
that photometric redshifts affect scales much larger that the
photometric redshift errors due to an effective reduction of the
number of Fourier modes and the smearing of the underlying true
clustering. 
\label{power} }
\end{center}
\end{figure*}

We investigate these effects directly on the measurement of the matter
power spectrum using large N-body simulations. We use the {\sc
l-basicc} ensemble of \citet{Angulo08}, which consists of
50 low-resolution, large volume simulations. Each has a volume of $2.4
({\rm pc}/h)^3$ and resolves halos more massive than $1\times
10^{13} \, {\rm M_\odot}/h$. The assumed cosmological parameters are
$\Omega_m=0.25$, $\Omega_{\Lambda}=0.75$, $h=0.73$, $n=1$ and
$\sigma_8 = 0.9$. Their huge volume makes the {\sc l-basicc}
simulations ideal to study the detectability of BAO in future
surveys. Photometric redshift errors are mimicked as a random
perturbation added to the particles' position along one direction
(line-of-sight). The perturbations are drawn from a Gaussian
distribution with various widths representing different degrees of
uncertainty in the photometric redshift. 
Despite their large volume, the {\sc l-basicc} boxes are more than an 
order of magnitude smaller than the volume which will be covered 
by the $3\pi$ survey. Hence, we present results for the {\it relative} 
change expected in the random errors for different photometric redshift 
errors. Angulo et~al. found that any systematic error in the recovery 
of the BAO scale was comparable to the sampling variance between 
{\sc l-basicc} realizations. To address the question of systematic 
errors we will need to use even larger volume simulations. Furthermore, 
new estimators are likely to be developed to extract the optimal BAO 
signal from photometric surveys. These more detailed questions are 
deferred to a later paper.

In the upper panels of Fig.~\ref{power} we show the mean, spherically
averaged power spectrum of the dark matter measured from the {\sc
l-basicc} simulations at $z=0.5$, along with its variance, in
photo-$z$ space (solid blue lines). The size of the photo-$z$ errors
are $\sigma_z = 0.01$ and $\sigma_z = 0.04$ (equivalent to $15.8$ and
$63.4$ $h^{-1}$Mpc at $z=0.5$) in the left- and right-hand panels
respectively. We have also plotted the power spectrum measured in
redshift-space (solid red lines) and the analytical expression of
Eq.~\ref{pk_conv} (dashed red line). By comparing the spectra in
redshift and photo-$z$ spaces, the additional damping described above
is evident. Also, we see that Eq.~\ref{pk_conv} describes
quantitatively this extra damping on scales where the power spectrum
is not shot-noise dominated.
 
In the lower panels of Fig.~\ref{power} we take a closer look at the
BAO by isolating them from the large-scale shape of the power
spectrum. We do this by dividing the power spectrum by a smoothed
version of the measurement. It is clear that since the number of
``noisy modes'' increases with the size of the photometric redshift
errors, the error on the power spectrum and therefore on the BAO also
increases. The visibility of the higher harmonic BAO is also reduced
as the photometric redshift error increases. In order to quantify the
loss of information, we have followed a standard technique to measure
BAO as described in \citet{Angulo08} (see also
\citealp{Percival07} and \citealp{Sanchez08}). The
method basically consists of dividing the measured power spectrum by a
smoothed version of the measurement. In this way, any long wavelength
gradient or distortion in the shape of the power spectrum is removed
which diminishes the impact of possible systematic errors due to
redshift space distortions, galaxy bias, nonlinear evolution and, in
the case described in this paper, photometric redshift
distortions. Then, we construct a model ratio using linear
perturbation theory, $P_{\rm lt}/P_{\rm smooth}$, which we fit to the
measured ratios. In the fitting procedure there are two free
parameters: (i) a damping factor to account for the destruction of BAO
peaks located at high $k$ by non-linear effects and redshift-space
distortions and (ii) a stretch factor, $\alpha$, which quantifies how
accurately we can measure the BAO wavelength. The latter gives a
simple estimate of how well we can constrain the dark energy equation
of state from BAO measurements alone.

\begin{figure}
\begin{center}
\includegraphics[width=8.5cm,angle=0]{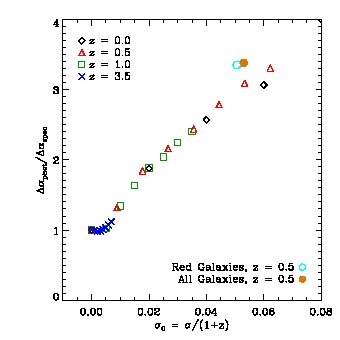}
\caption{The ratio of the error on the measurement of the BAO scale in
photo-$z$ space to that in redshift space (i.e. from a perfect
spectroscopic redshift) as a function of the magnitude of the
photometric redshift error. Assuming that the error on the measurement
scales with the square root of the volume, then the $y$-axis gives the
square root of the ratio of volumes of photometric to spectroscopic
surveys which achieve the same accuracy in the measurement of the BAO
scale. Note that this quantity is independent of the redshift at which
the measurement is made, i.e. it is independent of the degree of
nonlinearity present in the dark matter distribution. The cyan and 
brown circles give the results from using the actual distribution of 
photometric redshift errors, while the others assume a Gaussian error 
distribution shown in Fig.~\ref{PhotozErr}. 
\label{bao_err}}
\end{center}
\end{figure}

Fig~\ref{bao_err} shows the results of applying our fitting procedure
to the {\sc l-basicc} ensemble at different redshifts. On the $x$-axis
we plot the size of the photometric redshift error divided by $(1+z)$,
whilst on the $y$-axis we plot the predicted error on $\alpha$ divided
by the error we infer for an ideal spectroscopic survey (i.e. from the
power spectrum in redshift-space). Since the error on $\alpha$ scales
with the error on the power spectrum and the latter is proportional to
the square root of the volume of the survey, the $y$-axis should be
roughly equal to the square root of the factor by which the volume of
a photometric redshift needs to be larger than the volume of a
spectroscopic survey to achieve the same accuracy.
 
Several authors have investigated the implications of photometric
redshift errors on the clustering measurements in general and on the
BAO in particular
\citep{Seo03, Amendola05, Dolney06,
Blake05}. Our analysis improves upon these studies in
several ways: (i) we have included photometric redshift errors
directly into an realistic distribution of objects; (i) by using
N-body simulations, our calculation takes into account the effects
introduced by nonlinear evolution, nonlinear redshift-space
distortions and shot noise; (iii) the use of 50 different simulations
enables a robust and realistic estimation of the errors on the power
spectrum measurements; (iv) we have investigated how our results change 
if we use the actual distribution of photometric redshift errors 
(the cyan and brown circles in Fig~\ref{bao_err}), instead 
of a Gaussian fit and we find only a small additional degradation. 

These improvements lead to predictions that are somewhat different
from previous ones. For example, for $\Delta z=0.03$, 
\citet{Blake05} predict a factor of $\sim$10 for the reduction
of the effective volume of a photometric survey. Here, as shown in
Fig.~\ref{bao_err}, we find a reduction which is a factor 2 times
smaller than this (i.e. a volume reduction factor of $\sim$5). The
main difference between our analyses is that
\citet{Blake05} use only modes larger than $k_{\rm{max}} =
2/\sigma_z$, arguing that wavelengths shorter than the size of the
photometric redshift errors contribute only noise.  In reality, there
is a smooth transition around $k_{\rm{max}}$, with signal coming from all
wavenumbers (with different weighting, of course). In addition, the
neglect of nonlinear evolution (which erases the BAO at high
wavenumbers) also contributes to \citet{Blake05}
overestimating the reduction in effective volume. These two effects
together explain the disagreement between our results. 
\section{Discussion and conclusions}

We have described a method for constructing mock galaxy catalogues
which are well suited to aid in the preparation, and eventually in the
interpretation of large photometric surveys. We applied our mock
catalogues specifically to the data that will shortly begin to be
collected with PS1, the first of the 4 telescopes planned for the
Pan-STARRS system.

Our mock catalogue building method relies on the use of two
complementary theoretical tools: cosmological N-body simulations and a
semi-analytic model of galaxy formation. For this study, we have
employed the Millennium N-body simulations of
\citet{Springel05} together with galaxy properties calculated 
using the {\sc galform} model with the physics described by
\citet{Bower06}. Although this model gives quite a good match to the
local galaxy luminosity function in the B- and K-band, we refined the
match by applying a small correction of 0.15~mag to the luminosities
of all galaxies, so that the agreement with the SDSS luminosity
function is excellent. Similarly, we applied a correction to the
predicted galaxy sizes as a function of redshift in order to match the
SDSS distribution of Petrosian half-light radii in the $r$-band. As a
simple test, we showed that our galaxy formation model agrees very
well with the $r$-band number counts in the SDSS Commissioning and
{\sc deep}2 data over a range of 12 magnitudes.

We adopt a similar magnitude system as the SDSS, based on the use of
Petrosian magnitudes and use these to calculate the expected magnitude
limits for extended objects in the two surveys that PS1 will
undertake, the ``3$\pi$'' survey and the MDS. We find that, after 3
years, the 3$\pi$ survey will have detected $\sim 2\times 10^8$
galaxies in all 5 photometric bands ($g, r, i, z$ and $y$), with a
peak in the redshift distribution of $\sim$0.5 and an extended tail
containing about 10 million galaxies with $z>0.9$. The MDS will detect
$\sim 2\times 10^7$ galaxies, the redshift distribution peaking 
at $z\sim 0.8$, with 0.5 million
galaxies lying at $z>2$. Of the 5 PS1 bands $y$ is the shallowest and
removing the requirement that a galaxy be detected in this band more
than doubles the total numbers in the sample.

We have used our mock catalogues to take a first look at the 
accuracy of photometric redshifts in the PS1 photometric system. 
Photometric redshifts can be readily estimated using the public
Hyper-$z$ code of \citet{Bozonella00}. With the PS1 $grizy$
bands alone it is possible, in principle, to achieve an accuracy in
the 3$\pi$ survey of $\Delta z/(1+z)\sim 0.06$ in the range
$0.25<z<0.8$. This could be reduced to $\sim$0.05 by adding $J$ and
$K$ photometry from the UKIDDS (LAS) and could be improved even further
with a hypothetical $U$ band survey to 23 mag, although the samples
become progressively smaller as these additional bands are
added. Cutting
at the relatively shallow depth of the $y$-band 
is important in achieving these errors. Going deeper than the $y$-band
data would increase the sample size substantially, but the 
errors would increase to
$\sim$0.075. There is therefore a balance to be struck between
reducing the sample size (by about a factor of 2) 
which increases the accuracy of all the photometry, allows 
the $y$-band to be used and has the combined effect of 
increasing the photometric redshift accuracy. For the MDS an
accuracy of $\Delta z/(1+z)\sim 0.05$ is achievable for $0.02<z<1.5$
using the PS1 bands alone, with similar fractional improvements as for
the 3$\pi$ survey by the inclusion of $U$ and near infrared bands.

A dramatic improvement in photometric redshift accuracy can be
achieved for samples containing only red galaxies. We have shown that
it should be possible to identify a red sample (i.e. with red
rest-frame $g-r$ colours) directly from the photometric data using the
best-fit Hyper-$z$ templates. These samples can still contain large
numbers of galaxies. For example, an accuracy of $\Delta z/(1+z) \sim
0.02-0.04$ may be achievable for $\sim$100~million red galaxies at
$0.4<z<1.1$ in the 3$\pi$ survey. Similarly, for the MDS, this sort of
accuracy could be achieved for $\sim$30~million galaxies at
$0.4<z<2$. These estimates are all based on the ``off-the-shelf''
Hyper-$z$ code, without any tuning of the code for the PS1 setup. 
We expect that further improvements should be possible
by refining the photometric redshift estimator and tailoring it
specifically to the PS1 bands. 

Our analysis is based on the use of the {\sc galform} semi-analytic 
galaxy formation model. Although this model gives a good match to a 
large range of observed galaxy properties, it is based on a number 
of approximations and has uncertain elements which could be relevant 
to the estimation of photometric redshifts. These include the effects 
of reddening, assumptions about the frequency and duration of bursts 
and the use of the \citet{Bruzual93} stellar population synthesis 
libraries which are the same as assumed in our implementation of Hyper-$z$. 
We note that the star formation histories predicted by the model are much 
more varied and have a richer structure than those assumed to construct 
the Hyper-$z$ templates, and that the treatment of dust extinction is very 
different in {\sc galform}.  
\citet{Abdalla08} carried out a similar study to ours and reached 
similar conclusions about the size of the photometric redshift 
errors and the usefulness of additional filters in the NIR or far-UV. 
This is encouraging as Abdalla et~al. used a completely different 
photometric redshift estimator, ANNz, an artificial neural network 
code written by \cite{Collister04}. Furthermore, instead of using 
a galaxy formation model to generate a mock catalogue, these authors 
used a mixture of empirical and theoretical techniques to produce a 
set of galaxies on which to test their estimator. 

One of the main applications of the PS1 3$\pi$ survey will be to the
determination of the scale of baryonic acoustic oscillations used to
constrain the properties of the dark energy. We have investigated how
uncertainties in the photometric redshifts will degrade the
determination of the BAO scale and, in particular, we have quantified
the factor by which the effective volume of a photometric survey is
reduced by these uncertainties. We find that, with the sorts of
photometric redshift uncertainties that we have estimated for a red
sample, PS1 will achieve the same accuracy as a spectroscopic galaxy
survey containing 1/5 as many galaxies. Unfortunately, spectroscopy
for 20 million galaxies at $z\sim 1$ is not likely to be feasible for
some time. PS1 should be able to provide competitive estimates of the
BAO scale in the next few years.

\section*{ACKNOWLEDGEMENT}

YC is supported by the Marie Curie Early Stage Training Host Fellowship 
ICCIPPP, which is funded by the European Commission. This work 
was supported in part by the STFC Rolling Grant to the ICC for
research into ``The growth of structure in the Universe.''  
REA is supported by a STFC/British Petroleum sponsored Dorothy Hodgkin 
postgraduate award. CMB is funded by a Royal Society University Research 
Fellowship. CSF acknowledges a Royal Society Wolfson Research Merit Award. 
We acknowledge helpful comments from Stef Phleps and the OPINAS-LSS group
and Dave Wilman which helped to improve the presentation of an earlier 
draft. We also acknowledge useful comments from Richard Bower, 
Ken Chambers, Alan Heavens, Bob Joseph, Peder Norberg, John Peacock and Tom Shanks.

%\bibliography{GalMock_v7}
%\bibliographystyle{mn2e}

\input{GalMock.bbl}
\end{document}

%% file: aas_journals.tex
\def\aj{{AJ}}                   % Astronomical Journal
\def\apj{{ApJ}}                 % Astrophysical Journal
\def\apjl{{ApJ}}                % Astrophysical Journal, Letters
\def\apjs{{ApJS}}               % Astrophysical Journal, Supplement
\def\aap{{A\&A}}                % Astronomy and Astrophysics
\def\mnras{{MNRAS}}             % Monthly Notices of the RAS
\def\prd{{Phys.~Rev.~D}}        % Physical Review D
\def\pasp{{PASP}}               % Publications of the ASP
\def\nat{{Nature}}              % Nature

%% file: GalMock.bbl
\begin{thebibliography}{}

\bibitem[\protect\citeauthoryear{{Abdalla}, {Amara}, {Capak} \& et
  al.}{{Abdalla} et~al.}{2008}]{Abdalla08}
{Abdalla} F.~B.,  {Amara} A.,  {Capak} P.,    et al. 2008, \mnras, 387, 969

\bibitem[\protect\citeauthoryear{{Almeida}, {Baugh} \& {Lacey}}{{Almeida}
  et~al.}{2007}]{Almeida07}
{Almeida} C.,  {Baugh} C.~M.,    {Lacey} C.~G.,  2007, \mnras, 376, 1711

\bibitem[\protect\citeauthoryear{{Amendola}, {Quercellini} \&
  {Giallongo}}{{Amendola} et~al.}{2005}]{Amendola05}
{Amendola} L.,  {Quercellini} C.,    {Giallongo} E.,  2005, \mnras, 357, 429

\bibitem[\protect\citeauthoryear{{Angulo}, {Baugh}, {Frenk} \&
  {Lacey}}{{Angulo} et~al.}{2008}]{Angulo08}
{Angulo} R.~E.,  {Baugh} C.~M.,  {Frenk} C.~S.,    {Lacey} C.~G.,  2008,
  \mnras, 383, 755

\bibitem[\protect\citeauthoryear{{Baugh}}{{Baugh}}{2006}]{Baugh06}
{Baugh} C.~M.,  2006, Reports of Progress in Physics, 69, 3101

\bibitem[\protect\citeauthoryear{{Baugh}, {Cole} \& {Frenk}}{{Baugh}
  et~al.}{1996}]{Baugh96}
{Baugh} C.~M.,  {Cole} S.,    {Frenk} C.~S.,  1996, \mnras, 283, 1361

\bibitem[\protect\citeauthoryear{{Baugh}, {Lacey}, {Frenk}, {Granato}, {Silva},
  {Bressan}, {Benson} \& {Cole}}{{Baugh} et~al.}{2005}]{Baugh05}
{Baugh} C.~M.,  {Lacey} C.~G.,  {Frenk} C.~S.,  {Granato} G.~L.,  {Silva} L.,
  {Bressan} A.,  {Benson} A.~J.,    {Cole} S.,  2005, \mnras, 356, 1191

\bibitem[\protect\citeauthoryear{{Bender}, {Appenzeller} \&
  {B{\"o}hm}}{{Bender} et~al.}{2001}]{Bender01}
{Bender} R.,  {Appenzeller} I.,    {B{\"o}hm} A.,  2001, in Cristiani S.,
  Renzini A., Williams R. E., eds, Deep Fields. Springer, Berlin, p.~96

\bibitem[\protect\citeauthoryear{{Ben{\'{\i}}tez}}{{Ben{\'{\i}}tez}}{2000}]{Be%
nitez00}
{Ben{\'{\i}}tez} N.,  2000, \apj, 536, 571

\bibitem[\protect\citeauthoryear{{Benson}, {Frenk}, {Baugh}, {Cole} \&
  {Lacey}}{{Benson} et~al.}{2003}]{Benson03}
{Benson} A.~J.,  {Frenk} C.~S.,  {Baugh} C.~M.,  {Cole} S.,    {Lacey} C.~G.,
  2003, \mnras, 343, 679

\bibitem[\protect\citeauthoryear{{Blaizot}, {Szapudi}, {Colombi},
  {Budav{\`a}ri}, {Bouchet}, {Devriendt}, {Guiderdoni}, {Pan} \&
  {Szalay}}{{Blaizot} et~al.}{2006}]{Blaizot06}
{Blaizot} J.,  {Szapudi} I.,  {Colombi} S.,  {Budav{\`a}ri} T.,  {Bouchet}
  F.~R.,  {Devriendt} J.~E.~G.,  {Guiderdoni} B.,  {Pan} J.,    {Szalay} A.,
  2006, \mnras, 369, 1009

\bibitem[\protect\citeauthoryear{{Blaizot}, {Wadadekar}, {Guiderdoni},
  {Colombi}, {Bertin}, {Bouchet}, {Devriendt} \& {Hatton}}{{Blaizot}
  et~al.}{2005}]{Blaizot05}
{Blaizot} J.,  {Wadadekar} Y.,  {Guiderdoni} B.,  {Colombi} S.~T.,  {Bertin}
  E.,  {Bouchet} F.~R.,  {Devriendt} J.~E.~G.,    {Hatton} S.,  2005, \mnras,
  360, 159

\bibitem[\protect\citeauthoryear{{Blake} \& {Bridle}}{{Blake} \&
  {Bridle}}{2005}]{Blake05}
{Blake} C.,  {Bridle} S.,  2005, \mnras, 363, 1329

\bibitem[\protect\citeauthoryear{{Blake} \& {Glazebrook}}{{Blake} \&
  {Glazebrook}}{2003}]{Blake03}
{Blake} C.,  {Glazebrook} K.,  2003, \apj, 594, 665

\bibitem[\protect\citeauthoryear{{Blanton}, {Hogg}, {Bahcall}, {Brinkmann},
  {Britton}, {Connolly}, {Csabai}, {Fukugita} \& et al.}{{Blanton}
  et~al.}{2003}]{Blanton03}
{Blanton} M.~R.,  {Hogg} D.~W.,  {Bahcall} N.~A.,  {Brinkmann} J.,  {Britton}
  M.,  {Connolly} A.~J.,  {Csabai} I.,  {Fukugita} M.,    et al. 2003, \apj,
  592, 819

\bibitem[\protect\citeauthoryear{{Bolzonella}, {Miralles} \&
  {Pell{\'o}}}{{Bolzonella} et~al.}{2000}]{Bozonella00}
{Bolzonella} M.,  {Miralles} J.-M.,    {Pell{\'o}} R.,  2000, \aap, 363, 476

\bibitem[\protect\citeauthoryear{{Boulade}, {Charlot}, {Abbon} \& et
  al.}{{Boulade} et~al.}{2003}]{Boulade03}
{Boulade} O.,  {Charlot} X.,  {Abbon} P.,    et al. 2003, in {Iye} M.,
  {Moorwood} A.~F.~M.,  eds, Instrument Design and Performance for
  Optical/Infrared Ground-based Telescopes. Edited by Iye, Masanori; Moorwood,
  Alan F. M. Proceedings of the SPIE, Volume 4841, pp. 72-81 (2003). Vol.~4841
  of Presented at the Society of Photo-Optical Instrumentation Engineers (SPIE)
  Conference, {MegaCam: the new Canada-France-Hawaii Telescope wide-field
  imaging camera}.
pp 72--81

\bibitem[\protect\citeauthoryear{{Bouwens} \& {Silk}}{{Bouwens} \&
  {Silk}}{2002}]{Bouwens02}
{Bouwens} R.,  {Silk} J.,  2002, \apj, 568, 522

\bibitem[\protect\citeauthoryear{{Bower}, {Benson}, {Malbon}, {Helly}, {Frenk},
  {Baugh}, {Cole} \& {Lacey}}{{Bower} et~al.}{2006}]{Bower06}
{Bower} R.~G.,  {Benson} A.~J.,  {Malbon} R.,  {Helly} J.~C.,  {Frenk} C.~S.,
  {Baugh} C.~M.,  {Cole} S.,    {Lacey} C.~G.,  2006, \mnras, 370, 645

\bibitem[\protect\citeauthoryear{{Brunner}, {Szalay} \& {Connolly}}{{Brunner}
  et~al.}{2000}]{Brunner00}
{Brunner} R.~J.,  {Szalay} A.~S.,    {Connolly} A.~J.,  2000, \apj, 541, 527

\bibitem[\protect\citeauthoryear{{Bruzual} \& {Charlot}}{{Bruzual} \&
  {Charlot}}{1993}]{Bruzual93}
{Bruzual} A.~G.,  {Charlot} S.,  1993, \apj, 405, 538

\bibitem[\protect\citeauthoryear{{Bruzual} \& {Charlot}}{{Bruzual} \&
  {Charlot}}{2003}]{Bruzual03}
{Bruzual} G.,  {Charlot} S.,  2003, \mnras, 344, 1000

\bibitem[\protect\citeauthoryear{{Chambers}}{{Chambers}}{2006}]{Chambers06}
{Chambers} K.~C.,  2006, Pan-STARRS Mission Concept Statement for PS1, 23000200

\bibitem[\protect\citeauthoryear{{Coil}, {Newman}, {Kaiser}, {Davis}, {Ma},
  {Kocevski} \& {Koo}}{{Coil} et~al.}{2004}]{Coil04}
{Coil} A.~L.,  {Newman} J.~A.,  {Kaiser} N.,  {Davis} M.,  {Ma} C.-P.,
  {Kocevski} D.~D.,    {Koo} D.~C.,  2004, \apj, 617, 765

\bibitem[\protect\citeauthoryear{{Cole}, {Hatton}, {Weinberg} \&
  {Frenk}}{{Cole} et~al.}{1998}]{Cole98}
{Cole} S.,  {Hatton} S.,  {Weinberg} D.~H.,    {Frenk} C.~S.,  1998, \mnras,
  300, 945

\bibitem[\protect\citeauthoryear{{Cole}, {Lacey}, {Baugh} \& {Frenk}}{{Cole}
  et~al.}{2000}]{Cole00}
{Cole} S.,  {Lacey} C.~G.,  {Baugh} C.~M.,    {Frenk} C.~S.,  2000, \mnras,
  319, 168

\bibitem[\protect\citeauthoryear{{Cole}, {Norberg}, {Baugh}, {Frenk},
  {Bland-Hawthorn}, {Bridges}, {Cannon}, {Dalton} \& et al.}{{Cole}
  et~al.}{2001}]{Cole01}
{Cole} S.,  {Norberg} P.,  {Baugh} C.~M.,  {Frenk} C.~S.,  {Bland-Hawthorn} J.,
   {Bridges} T.,  {Cannon} R.,  {Dalton} G.,    et al. 2001, \mnras, 326, 255

\bibitem[\protect\citeauthoryear{{Cole}, {Percival}, {Peacock} \& et
  al.}{{Cole} et~al.}{2005}]{Cole05}
{Cole} S.,  {Percival} W.~J.,  {Peacock} J.~A.,    et al. 2005, \mnras, 362,
  505

\bibitem[\protect\citeauthoryear{{Coleman}, {Wu} \& {Weedman}}{{Coleman}
  et~al.}{1980}]{Coleman80}
{Coleman} G.~D.,  {Wu} C.-C.,    {Weedman} D.~W.,  1980, \apjs, 43, 393

\bibitem[\protect\citeauthoryear{{Colless}, {Dalton}, {Maddox} \& et
  al.}{{Colless} et~al.}{2001}]{Colless01}
{Colless} M.,  {Dalton} G.,  {Maddox} S.,    et al. 2001, \mnras, 328, 1039

\bibitem[\protect\citeauthoryear{{Collister} \& {Lahav}}{{Collister} \&
  {Lahav}}{2004}]{Collister04}
{Collister} A.~A.,  {Lahav} O.,  2004, \pasp, 116, 345

\bibitem[\protect\citeauthoryear{{Connolly}, {Csabai}, {Szalay}, {Koo}, {Kron}
  \& {Munn}}{{Connolly} et~al.}{1995}]{Connolly95}
{Connolly} A.~J.,  {Csabai} I.,  {Szalay} A.~S.,  {Koo} D.~C.,  {Kron} R.~G.,
   {Munn} J.~A.,  1995, \aj, 110, 2655

\bibitem[\protect\citeauthoryear{{Croton}, {Springel}, {White} \& et
  al.}{{Croton} et~al.}{2006}]{Croton06}
{Croton} D.~J.,  {Springel} V.,  {White} S.~D.~M.,    et al. 2006, \mnras, 365,
  11

\bibitem[\protect\citeauthoryear{{Csabai}, {Budav{\'a}ri}, {Connolly},
  {Szalay}, {Gy{\H o}ry}, {Ben{\'{\i}}tez}, {Annis}, {Brinkmann} \& et
  al.}{{Csabai} et~al.}{2003}]{Csabai03}
{Csabai} I.,  {Budav{\'a}ri} T.,  {Connolly} A.~J.,  {Szalay} A.~S.,  {Gy{\H
  o}ry} Z.,  {Ben{\'{\i}}tez} N.,  {Annis} J.,  {Brinkmann} J.,    et al. 2003,
  \aj, 125, 580

\bibitem[\protect\citeauthoryear{{Davis}, {Efstathiou}, {Frenk} \&
  {White}}{{Davis} et~al.}{1985}]{Davis85}
{Davis} M.,  {Efstathiou} G.,  {Frenk} C.~S.,    {White} S.~D.~M.,  1985, \apj,
  292, 371

\bibitem[\protect\citeauthoryear{{De Lucia}, {Springel}, {White}, {Croton} \&
  {Kauffmann}}{{De Lucia} et~al.}{2006}]{deLucia06}
{De Lucia} G.,  {Springel} V.,  {White} S.~D.~M.,  {Croton} D.,    {Kauffmann}
  G.,  2006, \mnras, 366, 499

\bibitem[\protect\citeauthoryear{{Dolney}, {Jain} \& {Takada}}{{Dolney}
  et~al.}{2006}]{Dolney06}
{Dolney} D.,  {Jain} B.,    {Takada} M.,  2006, \mnras, 366, 884

\bibitem[\protect\citeauthoryear{{Drory}, {Bender}, {Feulner}, {Hopp},
  {Maraston}, {Snigula} \& {Hill}}{{Drory} et~al.}{2003}]{Drory03}
{Drory} N.,  {Bender} R.,  {Feulner} G.,  {Hopp} U.,  {Maraston} C.,  {Snigula}
  J.,    {Hill} G.~J.,  2003, \apj, 595, 698

\bibitem[\protect\citeauthoryear{{Efstathiou}, {Moody}, {Peacock} \& et
  al.}{{Efstathiou} et~al.}{2002}]{Efstathiou02}
{Efstathiou} G.,  {Moody} S.,  {Peacock} J.~A.,    et al. 2002, \mnras, 330,
  L29

\bibitem[\protect\citeauthoryear{{Eisenstein}, {Zehavi}, {Hogg} \& et
  al.}{{Eisenstein} et~al.}{2005}]{Eisenstein05}
{Eisenstein} D.~J.,  {Zehavi} I.,  {Hogg} D.~W.,    et al. 2005, \apj, 633, 560

\bibitem[\protect\citeauthoryear{{Firth}, {Lahav} \& {Somerville}}{{Firth}
  et~al.}{2003}]{Firth03}
{Firth} A.~E.,  {Lahav} O.,    {Somerville} R.~S.,  2003, \mnras, 339, 1195

\bibitem[\protect\citeauthoryear{{Fukugita}, {Yasuda}, {Brinkmann}, {Gunn},
  {Ivezi{\'c}}, {Knapp}, {Lupton} \& {Schneider}}{{Fukugita}
  et~al.}{2004}]{Fukugita04}
{Fukugita} M.,  {Yasuda} N.,  {Brinkmann} J.,  {Gunn} J.~E.,  {Ivezi{\'c}} {\v
  Z}.,  {Knapp} G.~R.,  {Lupton} R.,    {Schneider} D.~P.,  2004, \aj, 127,
  3155

\bibitem[\protect\citeauthoryear{{Gabasch}, {Bender}, {Seitz} \& et
  al.}{{Gabasch} et~al.}{2004}]{Gabasch04}
{Gabasch} A.,  {Bender} R.,  {Seitz} S.,    et al. 2004, \aap, 421, 41

\bibitem[\protect\citeauthoryear{{Giallongo}, {D'Odorico}, {Fontana},
  {Cristiani}, {Egami}, {Hu} \& {McMahon}}{{Giallongo}
  et~al.}{1998}]{Giallongo98}
{Giallongo} E.,  {D'Odorico} S.,  {Fontana} A.,  {Cristiani} S.,  {Egami} E.,
  {Hu} E.,    {McMahon} R.~G.,  1998, \aj, 115, 2169

\bibitem[\protect\citeauthoryear{{Granato}, {Lacey}, {Silva} \& et
  al.}{{Granato} et~al.}{2000}]{Granato00}
{Granato} G.~L.,  {Lacey} C.~G.,  {Silva} L.,    et al. 2000, \apj, 542, 710

\bibitem[\protect\citeauthoryear{{Hatton}, {Devriendt}, {Ninin}, {Bouchet},
  {Guiderdoni} \& {Vibert}}{{Hatton} et~al.}{2003}]{Hatton03}
{Hatton} S.,  {Devriendt} J.~E.~G.,  {Ninin} S.,  {Bouchet} F.~R.,
  {Guiderdoni} B.,    {Vibert} D.,  2003, \mnras, 343, 75

\bibitem[\protect\citeauthoryear{{Helly}, {Cole}, {Frenk}, {Baugh}, {Benson},
  {Lacey} \& {Pearce}}{{Helly} et~al.}{2003}]{Helly03}
{Helly} J.~C.,  {Cole} S.,  {Frenk} C.~S.,  {Baugh} C.~M.,  {Benson} A.,
  {Lacey} C.,    {Pearce} F.~R.,  2003, \mnras, 338, 913

\bibitem[\protect\citeauthoryear{{Hewett}, {Warren}, {Leggett} \&
  {Hodgkin}}{{Hewett} et~al.}{2006}]{Hewett06}
{Hewett} P.~C.,  {Warren} S.~J.,  {Leggett} S.~K.,    {Hodgkin} S.~T.,  2006,
  \mnras, 367, 454

\bibitem[\protect\citeauthoryear{{Huang}, {Glazebrook}, {Cowie} \&
  {Tinney}}{{Huang} et~al.}{2003}]{Huang03}
{Huang} J.-S.,  {Glazebrook} K.,  {Cowie} L.~L.,    {Tinney} C.,  2003, \apj,
  584, 203

\bibitem[\protect\citeauthoryear{{Kang}, {Jing}, {Mo} \& {B{\"o}rner}}{{Kang}
  et~al.}{2005}]{Kang05}
{Kang} X.,  {Jing} Y.~P.,  {Mo} H.~J.,    {B{\"o}rner} G.,  2005, \apj, 631, 21

\bibitem[\protect\citeauthoryear{{Kang}, {Jing} \& {Silk}}{{Kang}
  et~al.}{2006}]{Kang06}
{Kang} X.,  {Jing} Y.~P.,    {Silk} J.,  2006, \apj, 648, 820

\bibitem[\protect\citeauthoryear{{Komatsu}, {Dunkley}, {Nolta} \& et
  al.}{{Komatsu} et~al.}{2008}]{Komatsu08}
{Komatsu} E.,  {Dunkley} J.,  {Nolta} M.~R.,    et al. 2008, ArXiv e-prints,
  0803.0547

\bibitem[\protect\citeauthoryear{{Lacey}, {Baugh}, {Frenk}, {Silva}, {Granato}
  \& {Bressan}}{{Lacey} et~al.}{2008}]{Lacey08}
{Lacey} C.~G.,  {Baugh} C.~M.,  {Frenk} C.~S.,  {Silva} L.,  {Granato} G.~L.,
   {Bressan} A.,  2008, \mnras, 385, 1155

\bibitem[\protect\citeauthoryear{{Lawrence}, {Warren}, {Almaini}, {Edge},
  {Hambly}, {Jameson}, {Lucas}, {Casali} \& et al.}{{Lawrence}
  et~al.}{2007}]{Lawrence07}
{Lawrence} A.,  {Warren} S.~J.,  {Almaini} O.,  {Edge} A.~C.,  {Hambly} N.~C.,
  {Jameson} R.~F.,  {Lucas} P.,  {Casali} M.,    et al. 2007, \mnras, 379, 1599

\bibitem[\protect\citeauthoryear{{Linder}}{{Linder}}{2003}]{Linder03}
{Linder} E.~V.,  2003, \prd, 68, 083504

\bibitem[\protect\citeauthoryear{{Malbon}, {Baugh}, {Frenk} \&
  {Lacey}}{{Malbon} et~al.}{2007}]{Malbon06}
{Malbon} R.~K.,  {Baugh} C.~M.,  {Frenk} C.~S.,    {Lacey} C.~G.,  2007,
  \mnras, 382, 1394

\bibitem[\protect\citeauthoryear{{Mobasher}, {Capak}, {Scoville} \& et
  al.}{{Mobasher} et~al.}{2007}]{Mobasher07}
{Mobasher} B.,  {Capak} P.,  {Scoville} N.~Z.,    et al. 2007, \apjs, 172, 117

\bibitem[\protect\citeauthoryear{{Norberg}, {Cole}, {Baugh}, {Frenk}, {Baldry},
  {Bland-Hawthorn}, {Bridges}, {Cannon} \& et al.}{{Norberg}
  et~al.}{2002}]{Norberg02}
{Norberg} P.,  {Cole} S.,  {Baugh} C.~M.,  {Frenk} C.~S.,  {Baldry} I.,
  {Bland-Hawthorn} J.,  {Bridges} T.,  {Cannon} R.,    et al. 2002, \mnras,
  336, 907

\bibitem[\protect\citeauthoryear{{Peacock} \& {Dodds}}{{Peacock} \&
  {Dodds}}{1994}]{Peacock94}
{Peacock} J.~A.,  {Dodds} S.~J.,  1994, \mnras, 267, 1020

\bibitem[\protect\citeauthoryear{{Percival}, {Baugh}, {Bland-Hawthorn} \& et
  al.}{{Percival} et~al.}{2001}]{Percival01}
{Percival} W.~J.,  {Baugh} C.~M.,  {Bland-Hawthorn} J.,    et al. 2001, \mnras,
  327, 1297

\bibitem[\protect\citeauthoryear{{Percival}, {Cole}, {Eisenstein}, {Nichol},
  {Peacock}, {Pope} \& {Szalay}}{{Percival} et~al.}{2007}]{Percival07}
{Percival} W.~J.,  {Cole} S.,  {Eisenstein} D.~J.,  {Nichol} R.~C.,  {Peacock}
  J.~A.,  {Pope} A.~C.,    {Szalay} A.~S.,  2007, \mnras, 381, 1053

\bibitem[\protect\citeauthoryear{{Petrosian}}{{Petrosian}}{1976}]{Petrosian76}
{Petrosian} V.,  1976, \apjl, 209, L1

\bibitem[\protect\citeauthoryear{{Pozzetti}, {Cimatti}, {Zamorani}, {Daddi},
  {Menci}, {Fontana}, {Renzini}, {Mignoli} \& et al.}{{Pozzetti}
  et~al.}{2003}]{Pozzetti03}
{Pozzetti} L.,  {Cimatti} A.,  {Zamorani} G.,  {Daddi} E.,  {Menci} N.,
  {Fontana} A.,  {Renzini} A.,  {Mignoli} M.,    et al. 2003, \aap, 402, 837

\bibitem[\protect\citeauthoryear{{Sanchez}, {Baugh} \& {Angulo}}{{Sanchez}
  et~al.}{2008}]{Sanchez08}
{Sanchez} A.~G.,  {Baugh} C.~M.,    {Angulo} R.,  2008, ArXiv e-prints, 804

\bibitem[\protect\citeauthoryear{{S{\'a}nchez}, {Baugh}, {Percival} \& et
  al.}{{S{\'a}nchez} et~al.}{2006}]{Sanchez06}
{S{\'a}nchez} A.~G.,  {Baugh} C.~M.,  {Percival} W.~J.,    et al. 2006, \mnras,
  366, 189

\bibitem[\protect\citeauthoryear{{Saunders}, {Frenk}, {Rowan-Robinson},
  {Lawrence} \& {Efstathiou}}{{Saunders} et~al.}{1991}]{Saunders91}
{Saunders} W.,  {Frenk} C.,  {Rowan-Robinson} M.,  {Lawrence} A.,
  {Efstathiou} G.,  1991, \nat, 349, 32

\bibitem[\protect\citeauthoryear{{Sawicki}, {Lin} \& {Yee}}{{Sawicki}
  et~al.}{1997}]{Sawicki97}
{Sawicki} M.~J.,  {Lin} H.,    {Yee} H.~K.~C.,  1997, \aj, 113, 1

\bibitem[\protect\citeauthoryear{{Seo} \& {Eisenstein}}{{Seo} \&
  {Eisenstein}}{2003}]{Seo03}
{Seo} H.-J.,  {Eisenstein} D.~J.,  2003, \apj, 598, 720

\bibitem[\protect\citeauthoryear{{Shen}, {Mo}, {White}, {Blanton}, {Kauffmann},
  {Voges}, {Brinkmann} \& {Csabai}}{{Shen} et~al.}{2003}]{Shen03}
{Shen} S.,  {Mo} H.~J.,  {White} S.~D.~M.,  {Blanton} M.~R.,  {Kauffmann} G.,
  {Voges} W.,  {Brinkmann} J.,    {Csabai} I.,  2003, \mnras, 343, 978

\bibitem[\protect\citeauthoryear{{Sowards-Emmerd}, {Smith}, {McKay}, {Sheldon},
  {Tucker} \& {Castander}}{{Sowards-Emmerd} et~al.}{2000}]{Sowards-Emmerd00}
{Sowards-Emmerd} D.,  {Smith} J.~A.,  {McKay} T.~A.,  {Sheldon} E.,  {Tucker}
  D.~L.,    {Castander} F.~J.,  2000, \aj, 119, 2598

\bibitem[\protect\citeauthoryear{{Springel}, {White}, {Jenkins}, {Frenk},
  {Yoshida}, {Gao}, {Navarro}, {Thacker}, {Croton}, {Helly}, {Peacock}, {Cole},
  {Thomas}, {Couchman}, {Evrard}, {Colberg} \& {Pearce}}{{Springel}
  et~al.}{2005}]{Springel05}
{Springel} V.,  {White} S.~D.~M.,  {Jenkins} A.,  {Frenk} C.~S.,  {Yoshida} N.,
   {Gao} L.,  {Navarro} J.,  {Thacker} R.,  {Croton} D.,  {Helly} J.,
  {Peacock} J.~A.,  {Cole} S.,  {Thomas} P.,  {Couchman} H.,  {Evrard} A.,
  {Colberg} J.,    {Pearce} F.,  2005, \nat, 435, 629

\bibitem[\protect\citeauthoryear{{Strauss}, {Weinberg}, {Lupton} \& et
  al.}{{Strauss} et~al.}{2002}]{Strauss02}
{Strauss} M.~A.,  {Weinberg} D.~H.,  {Lupton} R.~H.,    et al. 2002, \aj, 124,
  1810

\bibitem[\protect\citeauthoryear{{Tegmark}, {Strauss}, {Blanton} \& et
  al.}{{Tegmark} et~al.}{2004}]{Tegmark04}
{Tegmark} M.,  {Strauss} M.~A.,  {Blanton} M.~R.,    et al. 2004, \prd, 69,
  103501

\bibitem[\protect\citeauthoryear{{Trujillo}, {F{\"o}rster Schreiber},
  {Rudnick}, {Barden}, {Franx}, {Rix}, {Caldwell}, {McIntosh} \& et
  al.}{{Trujillo} et~al.}{2006}]{Trujillo06}
{Trujillo} I.,  {F{\"o}rster Schreiber} N.~M.,  {Rudnick} G.,  {Barden} M.,
  {Franx} M.,  {Rix} H.-W.,  {Caldwell} J.~A.~R.,  {McIntosh} D.~H.,    et al.
  2006, \apj, 650, 18

\bibitem[\protect\citeauthoryear{{Tyson}}{{Tyson}}{2002}]{Tyson02}
{Tyson} J.~A.,  2002, in {Tyson} J.~A.,  {Wolff} S.,  eds, Survey and Other
  Telescope Technologies and Discoveries. Edited by Tyson, J. Anthony; Wolff,
  Sidney. Proceedings of the SPIE, Volume 4836, pp. 10-20 (2002). Vol.~4836 of
  Presented at the Society of Photo-Optical Instrumentation Engineers (SPIE)
  Conference, {Large Synoptic Survey Telescope: Overview}.
pp 10--20

\bibitem[\protect\citeauthoryear{{White} \& {Frenk}}{{White} \&
  {Frenk}}{1991}]{White91}
{White} S.~D.~M.,  {Frenk} C.~S.,  1991, \apj, 379, 52

\bibitem[\protect\citeauthoryear{{White}, {Tully} \& {Davis}}{{White}
  et~al.}{1988}]{White88}
{White} S.~D.~M.,  {Tully} R.~B.,    {Davis} M.,  1988, \apjl, 333, L45

\bibitem[\protect\citeauthoryear{{Yasuda}, {Fukugita}, {Narayanan}, {Lupton},
  {Strateva}, {Strauss}, {Ivezi{\'c}}, {Kim} \& et al.}{{Yasuda}
  et~al.}{2001}]{Yasuda01}
{Yasuda} N.,  {Fukugita} M.,  {Narayanan} V.~K.,  {Lupton} R.~H.,  {Strateva}
  I.,  {Strauss} M.~A.,  {Ivezi{\'c}} {\v Z}.,  {Kim} R.~S.~J.,    et al. 2001,
  \aj, 122, 1104

\bibitem[\protect\citeauthoryear{{Yasuda}, {Fukugita} \& {Schneider}}{{Yasuda}
  et~al.}{2007}]{Yasuda07}
{Yasuda} N.,  {Fukugita} M.,    {Schneider} D.~P.,  2007, \aj, 134, 698

\bibitem[\protect\citeauthoryear{{York}, {Adelman}, {Anderson} Jr. \& et
  al.}{{York} et~al.}{2000}]{York00}
{York} D.~G.,  {Adelman} J.,  {Anderson} Jr. J.~E.,    et al. 2000, \aj, 120,
  1579

\bibitem[\protect\citeauthoryear{{Yoshida}, {Stoehr}, {Springel} \&
  {White}}{{Yoshida} et~al.}{2002}]{Yoshida02}
{Yoshida} N.,  {Stoehr} F.,  {Springel} V.,    {White} S.~D.~M.,  2002, \mnras,
  335, 762

\end{thebibliography}
